\documentclass[namedreferences]{solarphysics}

\usepackage[hyperref,optionalrh]{spr-sola-addons}
\usepackage{graphicx}        % For eps figures, newer & more powerfull
\usepackage{color}           % For color text: \color command
\usepackage{breakurl}        % For breaking URLs easily trough lines
\chardef\us=`\_

\begin{document}

\begin{article}
\begin{opening}

\title{A model of a tidally synchronized solar dynamo}

%\author[addressref={},corref,email={F.Stefani@hzdr.de}]{\inits{F.}\fnm{F.}~\lnm{Stefani}}%\sep
%\author[addressref=aff1,email={e-mail.b@mail.com}]{\inits{F.}\fnm{Fisrt}~\lnm{Author-b}}%\sep
\author[corref,email={F.Stefani@hzdr.de}]{\inits{F.}\fnm{F.}~\lnm{Stefani}}%\sep
\author{\inits{A.}\fnm{A.}~\lnm{Giesecke}}
\author{\inits{T.}\fnm{T.}~\lnm{Weier}}%\sep
\address{Helmholtz-Zentrum Dresden -- Rossendorf, Bautzner Landstr. 400,
D-01328 Dresden, Germany}

\runningauthor{F. Stefani {\it et al.}}
\runningtitle{A model of a tidally synchronized solar dynamo}

\begin{abstract} We discuss a solar dynamo model of
Tayler-Spruit type whose $\Omega$-effect 
is conventionally produced by a solar-like 
differential rotation 
but whose $\alpha$-effect is assumed to be
periodically modulated by planetary tidal forcing. 
This resonance-like effect has its rationale 
in the tendency of the current-driven Tayler 
instability to undergo
intrinsic helicity oscillations which, in turn, 
can be synchronized by periodic 
tidal perturbations.
Specifically, we focus on the  11.07 years
alignment periodicity of the 
tidally dominant planets Venus, Earth, and Jupiter,
whose persistent synchronization with the solar dynamo is
briefly touched upon.  
The typically emerging  dynamo modes are dipolar
fields, oscillating with a 22.14 years period
or pulsating with an 11.07 years period, 
but also quadrupolar fields with corresponding 
periodicities.
In the absence of any constant part of 
$\alpha$, we prove the subcritical
nature of this Tayler-Spruit type dynamo.
The resulting amplitude of the $\alpha$ oscillation  
that is required for dynamo action
turns out to lie 
in the order of 1 m/s, which seems not implausible for the
sun. When starting with a more classical, non-periodic 
part of $\alpha$, even less of the
oscillatory $\alpha$ part is needed 
to synchronize the entire dynamo.
Typically, the dipole solutions 
show butterfly diagrams, although 
their shapes are not convincing yet.
Phase coherent transitions between dipoles and quadrupoles,
which are reminiscent of the 
observed behaviour during the
Maunder minimum, can be easily triggered 
by long-term variations of dynamo parameters, 
but may also
occur spontaneously even for fixed parameters. 
Further interesting features 
of the model are the typical second intensity peak 
and the intermittent
appearance of reversed helicities in both 
hemispheres. 

\end{abstract}
\end{opening}
%-------------------------------------------------
\section{Introduction}

Asking ``Is there a chronometer hidden deep 
in the sun?'', \cite{Dicke1978} had analyzed the
similarity of the solar cycle 
with either a random walk process or, alternatively,
a  clocked process being perturbed by random
fluctuations. While his statistical results pointed
in favour of a clocked process, with 
shorter cycles usually being followed by 
longer ones as if the Sun remembered the 
correct phase, his conclusion
was later criticized by \cite{Gough1990} 
and \cite{Hoyng1996} as relying
on a too short time series of just 25 cycles.      

The closely related discussion, initiated by  
\cite{Wolf1859} and later entered
by \cite{Bollinger1952,Takahashi1968,Wood1972,Opik1972,CondonSchmidt1975,Grandpierre1996,Palus2000,Hung2007,Wilson2013,Okhlopkov2014,Poluianov2014},
of whether the Hale cycle of the Sun is synchronized 
with the  11.07 years alignment 
cycle of the tidally dominant planets 
Venus, Earth and Jupiter, 
was recently fueled by
\cite{Okhlopkov2016}
who demonstrated an amazing 
parallelism of both time series 
for the last 1000 years.

\begin{figure}[h!]
\includegraphics[width=120mm]{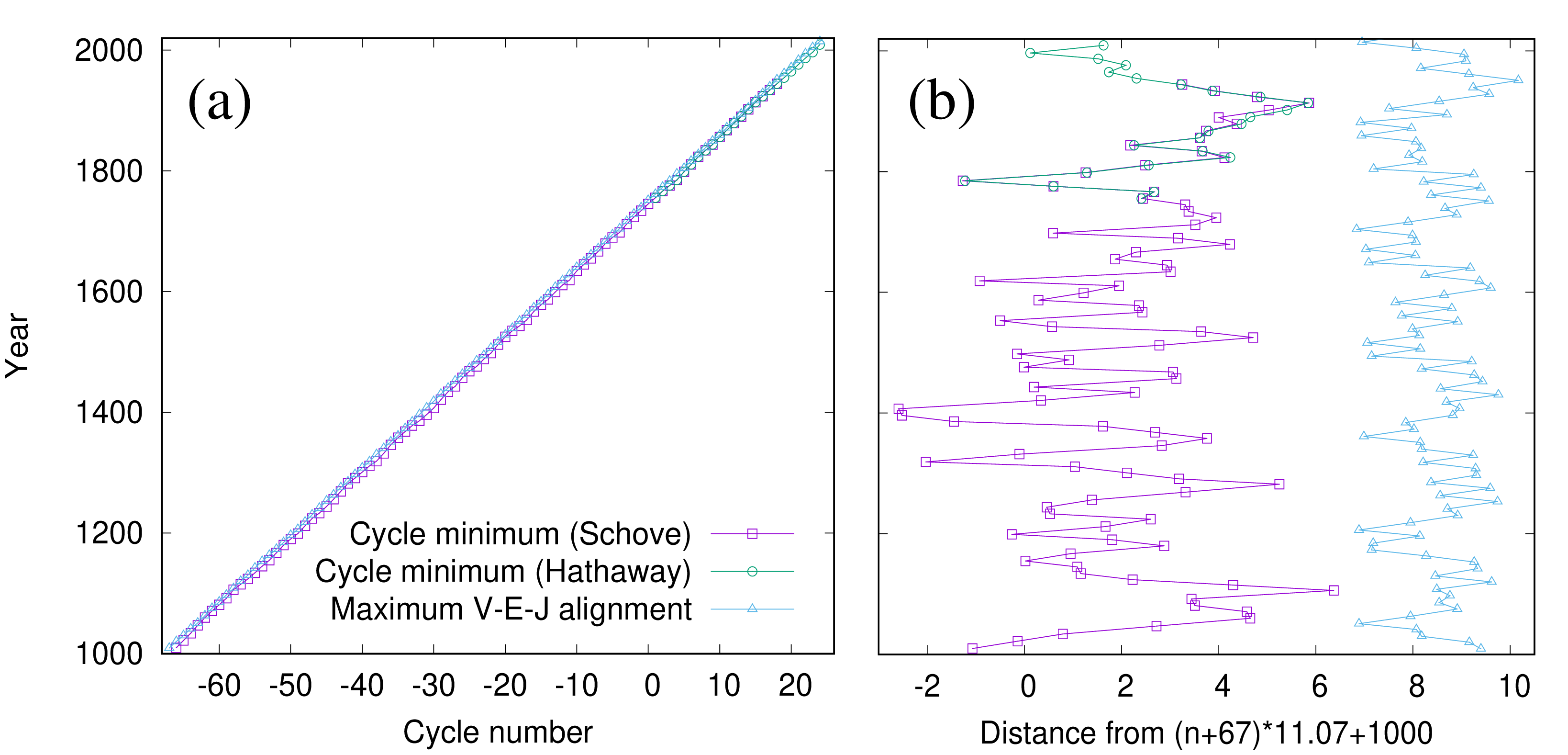}
\caption{(a) Time series of the minima of the
solar cycle according to
\cite{Schove1955,Schove1983} and \cite{Hathaway2010}, 
and of the maximum alignment of the 
 Venus-Earth-Jupiter system. (b) Deviation of 
 the time series 
from a linear function  $f(n)=11.07 (n+67)+1000$ 
of the cycle number $n$.
}
\label{Fig:vergleich}
\end{figure}

In Figure \ref{Fig:vergleich}(a) we illustrate
the sequence $t_n$ of the solar minima, as taken 
from  \cite{Schove1955,Schove1983} and \cite{Hathaway2010}, 
together with the corresponding sequence of 
the maximum Venus-Earth-Jupiter alignments
according to 
\cite{Okhlopkov2016}, which we have 
re-calculated and confirmed for the last 1000 years. 
In Figure \ref{Fig:vergleich}(b)
we show in detail the deviations (or residuals) 
$\delta_n=t_n-((n+67)\times 11.07+1000)$ of the
three time series from a
linear function of the cycle number $n$, with
a presumed cycle duration of 11.07 years. Note the
persistent closeness of the
solar cycle to this linear curve, which 
does not even change during the Maunder 
\citep{Beer1998} and Sp\"orer \citep{Miyahara2006} 
minima. While there is no evidence for any 
{\it local clocking} between the maximum Venus-Earth-Jupiter 
alignments and the cycle duration, both data sets remain 
{\it globally clocked}, with the 
solar minima instants never leaving a $\pm 4.5$ years  
band around the linear trend with 11.07 years period.

\begin{figure}[h!]
\begin{center}
\includegraphics[width=90mm]{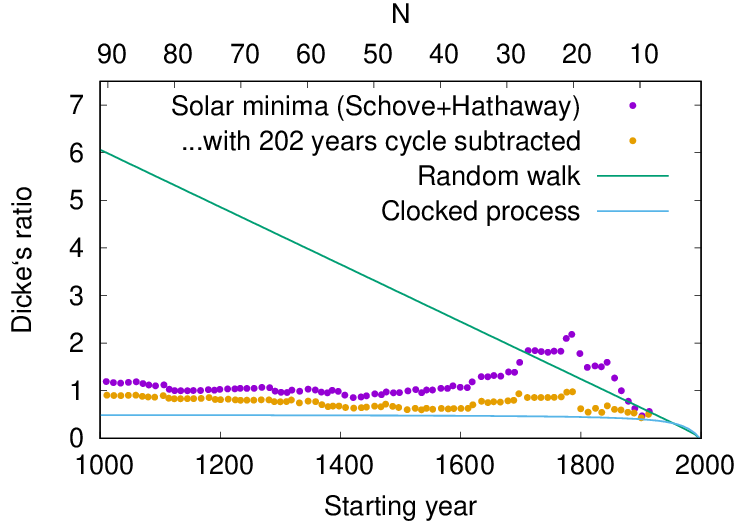}
\end{center}
\caption{Dicke's ratio between the mean square of the 
residuals to the mean square of the differences of two
subsequent residuals in dependence on the
number $N$ of cycles taken into account, 
for a random walk process (green line, converging towards $N/15$), a
clocked process (blue line, converging towards 0.5), and 
the real solar cycle minima data (violet dots) 
from Schove and Hathaway.
Despite the deterioration of the data's reliableness and
accuracy for the time before 1600, a tendency towards
a clocked process can still be observed. 
This becomes even more pronounced (orange dots) when a 
data-fitted
Suess-de Vries type cycle (yielding a period of 202 years)
is subtracted from the data.
}
\label{Fig:dicke}
\end{figure}

If we take those data of Schove and Hathaway (with all
due caveats regarding their reliableness and accuracy 
before the year ~1600, say), we can recompute 
Dicke's ratio 
$\sum_{i=n}^{24} \delta^2_i/\sum_{i=n}^{24} (\delta_i-\delta_{i-1})^2$ 
of the 
mean square of the residuals $\delta_i$ to the mean square of the
differences $\delta_i-\delta_{i-1}$ 
between two subsequent residuals. As stated by
Dicke, the dependence of this ratio on the number $N=24-n+1$ 
of cycles 
taken into account 
reads $(N+1)(N^2-1)/(3(5 N^2+6N-3))$ for a random walk 
process and $(N^2-1)/(2(N^2+2N+3))$
for a clocked process. Hence, for $N \rightarrow \infty$, 
the random 
walk ratio converges towards $N/15$, while the 
clocked process ratio converges to 0.5.        
Both curves are shown in  Figure \ref{Fig:dicke}, together with
Dicke's ratio  computed for the actual Schove/Hathaway data
(violet dots). 
While Dicke's original database 
was restricted to 25 cycles starting approximately at 1705, which 
made it hard to draw a solid conclusion about the character of
the process, the enlarged database of Schove indicates 
that the real process proceeds (for large $N$) 
much closer
to a clocked process than to a random walk process.

This feature becomes even  more pronounced 
(orange dots) if we first subtract 
from the data a noticable Suess-de Vries type cycle 
(yielding a period of 202 years when fitted to the Schove/Hathaway 
data
between the years 1000 and 2009).
Then, the overshoot peak around the year 1800 is strongly 
reduced. Down to the year 1400 a further 
convergence towards the ultimate value 0.5 could 
even be expected, and the slight increase prior to this year 
might be guessed to be caused by the 
deteriorating accuracy of a data. At any rate, 
it seems worthwhile
(but goes far beyond the scope of this paper)
to look for better quality data for those early 
years, and to study the influence of the
long-term cycles (Gleissberg, Suess-de Vries, 
Eddy) in a more systematic manner.

However impressive that coincidence of the
solar cycle with a clocked process in general, and 
with the maximum Venus-Earth-Jupiter 
alignments in particular, may look like: the 
counter-arguments against any sort of external
synchronization are serious as well. Indeed, 
the typical tidal 
acceleration of those planets (in the order of 
10$^{-10}$\,m/s$^2$) is 
tiny compared to other 
acceleration terms in the sun 
\citep{CondonSchmidt1975,DeJager2005,Callebaut2012}.
Even if  the typical tidal height of
$h_{\rm tidal} =
G m R^2_{\rm tacho}/(g_{\rm tacho} d^3) \approx 1$\,mm 
(exerted by a planet of mass $m$ at distance $d$ from the Sun) 
could be fully translated  
into a corresponding velocity
of $v \sim (2 g_{\rm tacho} h_{\rm tidal})^{1/2} \approx 1$\,m/s
(employing the huge gravity at the tachocline of 
$g_{\rm tacho} \approx 500$\,m/s$^2$ \citep{Wood2010}),
a physically realistic synchronization mechanism
based on these tides is still hardly conceivable.
 
Although the competing forces in the 
convection zone are prohibitively large for any planetary 
synchronization mechanism to get a chance to work there, 
things may be more subtle in the stably
stratified  tachocline region. A promising idea about a 
putative planetary influence, as first 
discussed by \cite{Abreu2012},
relies on periodic tidal perturbations
of the adiabaticity in the tachocline region, 
whose value is decisive for its
storage capacity for magnetic flux tubes. While primarily 
discussed with view 
on long-term modulations of the solar dynamo, there 
is no prior reason not to apply the same concept to the 
basic Hale cycle as well. 
In a recent paper \citep{Stefani2018} we  made a first 
attempt to validate this idea in the framework of a
simplified Babcock-Leighton type model,
employing the time-delay concept of 
\cite{Wilmotsmith2006}. 
Specifically, the tidal perturbations
were emulated as periodic changes of the 
critical magnetic field strength beyond which 
flux tubes  would erupt to the solar surface. 
Although our results, obtained 
in a limited parameter region, were essentially negative, 
we still consider this synchronization 
mechanism as rather attractive, and would 
like to encourage further work in this 
direction.

Yet another promising synchronization mechanism was
first delineated  by \cite{Weber2015} 
and later corroborated
in detail by \cite{Stefani2016,Stefani2018}.
It starts from the  numerical 
observation that the current-driven, kink-type
Tayler instability (TI) 
\citep{Tayler1973,Pittstayler1985,Gellert2011,Seilmayer2012,Ruediger2013,Ruediger2015,Stefani2015} 
has an intrinsic tendency for 
oscillations of the helicity and 
the $\alpha$-effect related to it.

At this point, a few general remarks on kink-type 
instabilities, 
and their applicability to stellar dynamo
models, may be appropriate: 
the notion Tayler-Spruit dynamo referred originally 
to the idea of \cite{Spruit2002}
who had proposed a non-linear, subcritical
dynamo in which the poloidal-to-toroidal field 
transformation is
conventionally accomplished by the $\Omega$-effect, 
while the toroidal-to-poloidal transformation
starts only when the toroidal field acquires sufficient strength 
to become  unstable to the non-axisymmetric,
current-driven  TI.
A fundamental flaw of this dynamo concept was revealed  
by \cite{Zahn2007}
who argued that any emerging 
non-axisymmetric ($m=1$) TI mode 
would be topologically unsuitable for regenerating the 
dominant axisymmetric ($m=0$) toroidal field. 
Fortunately, the same authors offered a possible remedy 
for the Tayler-Spruit dynamo concept provided 
that the $m=1$ TI would produce an 
$\alpha$-effect (comprising some $m=0$ component).
In hindsight, it appears 
that this idea had been investigated  more 
than a decade earlier by \cite{Ferrizmas1994}. 
Working in the flux-tube approximation, 
these authors had derived the $\alpha$-effect 
connected with the 
kink-instability and pointed out its crucial 
importance for closing the dynamo loop.

Beyond flux-tube approximation, 
the existence of any TI-related $\alpha$-effect 
is still a matter of debate.
Various authors \citep{Chatterjee2011,Gellert2011,Bonanno2012,Bonanno2017} 
had evidenced spontaneous symmetry breaking between 
left- and right-handed TI modes, leading
indeed to a finite value of $\alpha$, but
mainly
for comparably large values of the magnetic 
Prandtl number [$Pm$], i.e., the ratio between
viscosity and magnetic diffusivity.
Things are different, though,  
for the case of low $Pm$, as it is typical for 
the solar tachocline region
where $Pm$ is believed to lie in the range 10$^{-3}$...10$^{-2}$. 
In this limit of small $Pm$, we had numerically 
observed (although in the simplified 
geometry of a full cylinder) 
a tendency of the TI to undergo
{\it oscillations} of the helicity and the $\alpha$-effect 
\citep{Weber2013,Weber2015}.
Remarkably, those oscillations between  
left- and right-handed $m=1$ TI modes do barely change
the energy content of the instability, which makes them
very susceptible to weak $m=2$ perturbations \citep{Stefani2016}. 
This fact may indeed be the key for the easy 
synchronizability of the $\alpha$-effect with the tiny  
tidal forces as exerted by 
planets.

The resonant reaction of $\alpha$ on tidal excitations 
was later incorporated into a very simple ordinary differential
equation (ODE) model of an
$\alpha-\Omega$ dynamo which turned out to produce 
oscillations with period doubling \citep{Stefani2016}.
In this way  it was argued that the 11.07 years 
tidal perturbations could lead to a resonant 
excitation of an 11.07 years oscillation of the 
TI-related $\alpha$-effect, and thereby to a 
22.14 year Hale cycle of the entire dynamo.

In  \cite{Stefani2018}, it was specified that
such field {\it oscillations} occur only 
in certain bands of the magnetic diffusion time 
$\tau$, while
for  intervening bands they were replaced by
field {\it pulsations} with 11.07 years period. 
Noteworthy was the persistent 
phase coherence 
when passing from oscillations to pulsations, 
and back.
What could not be resolved by this simple 
ODE system 
(despite some progress in \cite{Stefani2017})
was the spatio-temporal specifics of the transitions
between oscillations and pulsations, for which 
higher dimensional modeling is definitely required.

As a  sequel to \cite{Stefani2016,Stefani2018},
the present paper investigates this
spatio-temporal behaviour of a tidally synchronized 
dynamo of the Tayler-Spruit type.
For that purpose, we replace  the ODE system 
by a partial differential
equation (PDE) system with the co-latitude as the only 
spatial variable. Similar radially averaged, pseudo-Cartesian 
models (although without any synchronization aspect) 
have been studied by many authors 
\citep{Parker1955,Schmalz1991,Jennings1991,Roald1997,Kuzanyan1997}, 
which will allow us, in Section 2 and the Appendix, to 
compare and validate our numerical method.

In Section 3, we will analyze in detail 
a synchronized, subcritical dynamo of
Tayler-Spruit type in its purest form. For that
purpose, we use a latitudinal
dependence of the  $\Omega$-effect as inferred from 
helioseismology \citep{Charbonneau1999}, and 
restrict the $\alpha$-effect to its
11.07 years periodic part whose amplitude has the 
same resonance-like dependence on the 
toroidal field as originally proposed in 
\cite{Stefani2016}. 
Since, for weak fields, this resonance term is proportional to 
the square of the field, it cannot yield a
linear instability. Instead, the dynamo needs some finite
field amplitude to start off. Apart from a detailed discussion
of the dependence of this sub-critical dynamo 
on the initial conditions, we will
also argue that the typical resulting amplitudes of $\alpha$
are on the order of 1 m/s, which seems not unrealistic
for the solar dynamo. Depending on some parameter choices, the 
arising fields are dipoles or quadrupoles, which can 
either oscillate with a 22.14 years period or pulsate with
an 11.07 years period. We also observe intermediate 
states between oscillations and
pulsations, which are reminiscent of the Gnevyshev-Ohl rule 
\citep{Gnevyshev}, which states that 
the sunspot numbers over an odd cycle exceeds that 
of the preceding even cycle.
During transitions between dipoles and quadrupoles, 
hemispherical dynamos are partly observed, too.

The oscillatory dipole solutions show, for 
high latitudes, poleward migration 
(``rush to the poles''), and for 
low latitudes a sort of butterfly diagram
in the correct direction, although its form is not 
completely convincing yet. 
Further interesting features to be discussed here
are a second intensity maximum, comparable 
to the double peak of the solar dynamo,
and the intermediate 
appearance of reversed helicities in the two hemispheres. 
The latter fact,
which is a direct consequence of the synchronized, oscillatory
character of $\alpha$, might be related
to the current-helicity observations of 
\cite{Zhang2010,Pipin2013}.

In section 4, we will soften the pure Tayler-Spruit principle 
by combining the periodic part
of $\alpha$ with a more standard, 
non-periodic term that is asymmetric with respect 
to the equator and only quenched by the toroidal field in 
the conventional manner. In the limiting case of a conventional
$\alpha-\Omega$ dynamo we obtain dipoles or quadrupoles with
typical periods between 20 and 40 years. When adding 
to this standard dynamo
our resonant periodic $\alpha$ term, we can 
easily enslave the  
dynamo to the 22.14 years periodicity, partly with some 
intermediate 2:3 synchronization to a 33.21 years period. 
Remarkably, the amplitude of the oscillatory part of 
$\alpha$ that
is required for this synchronization turns out to be
significantly smaller (below 1 m/s) 
than the typical values needed  for the 
purely non-linear dynamo as discussed above. 
By increasing the oscillatory part of $\alpha$ we obtain then 
a sequence of oscillatory quadrupoles, hemispherical dynamos,
dipoles with a strong Gnevyshev-Ohl tendency, and regularly 
oscillating dipoles.
When adding some noise to the non-periodic $\alpha$ term, 
the conventional $\alpha-\Omega$ model
and the synchronized ``hybrid'' model exhibit typical 
features of a random walk process and a clocked 
process, respectively, as will be illustrated by 
Dicke's ratio.

In section 5, we show how long-term changes of various
dynamo parameters (e.g., the portion of the 
periodic $\alpha$ part or the term which governs field 
losses by magnetic buoyancy)
are capable of producing transitions 
between dipole and quadrupole fields, a behaviour 
for which some observational evidence exists from 
the Maunder minimum \citep{Sokoloff1994,Arlt2009,Moss2017,Weiss2016}. 
A robust feature  of our synchronization model 
is the phase coherence which is maintained throughout 
such transitions.

The paper closes with a summary, 
a discussion of open questions, 
including the applicability of the general idea 
to other $m=1$ instabilities or flow structures, 
in particular the recently discussed magneto-Rossby waves of
the tachocline
\citep{McIntosh2017,Dikpati2017,Zaqarashvili2018},
and a call for higher dimensional simulations of this type of  
tidally synchronized solar dynamo model.

\section{The numerical model}

In this section we set-up the dynamo model and discuss its numerical
implementation.
We work with a system of partial differential equations, whose
spatial variable is restricted to the solar co-latitude
$\theta$. While similar models have been utilized by a number of
authors \citep{Parker1955,Schmalz1991,Jennings1991,Roald1997,Kuzanyan1997}, 
we use the specific formulation of \cite{Jennings1991}.

We focus on the axisymmetric magnetic field which is split into a
poloidal component ${\bf{B}}_P=\nabla \times (A {\bf{e}}_{\phi})$
and a toroidal component ${\bf{B}}_T=B {\bf{e}}_{\phi}$.
Introducing the helical turbulence parameter $\alpha$ and
the radial derivative $\omega=\sin(\theta) d (\Omega r)/dr$ of the 
azimuthal velocity, we arrive at the
one-dimensional  $\alpha - \Omega$ dynamo model
\begin{eqnarray} 
  \frac{{\partial} B(\theta,t)}{{\partial} t} &=& \omega(\theta,t) \frac{\partial A(\theta,t)}{\partial \theta} 
  + \frac{\partial^2 B(\theta,t)}{\partial \theta^2} -\kappa B^3(\theta,t) \\
    \frac{{\partial} A(\theta,t) }{{\partial} t} &=& \alpha(\theta,t) B(\theta,t) 
    + \frac{\partial^2 A(\theta,t)}{\partial \theta^2} , 
    \label{system_tayler}
   \end{eqnarray}
wherein $A(\theta,t)$ represents the vector potential of the 
poloidal field at co-latitude $\theta$ (running between 0 and $\pi$) 
and time $t$, and 
$B(\theta,t)$ the corresponding toroidal field. 
Here, $\alpha$ and $\omega$ 
denote the non-dimensionalized
versions of the dimensional quantities $\alpha_{\rm dim}$ and 
$\omega_{\rm dim}$,
according to $\alpha=\alpha_{\rm dim} R/\eta$  and
$\omega=\omega_{\rm dim} R^2/\eta$, where $R$ is the radius of the
considered dynamo region (we will later use here the radius of the tachocline) 
and $\eta$ is the magnetic diffusivity which is connected with the 
conductivity $\sigma$ via $\eta=1/(\mu_o \sigma)$. The time is non-dimensionalized by
the diffusion time, i.e. $t=t_{\rm dim} \eta/R^2$.

The not so familiar term 
$\kappa B^3(\theta,t)$, as introduced by \cite{Jones1983,Jennings1991}, 
has been included to account 
for losses owing to magnetic buoyancy, on the assumption that the 
escape velocity is 
proportional to $B^2$. While this term is not essential 
for our synchronization model, it may provide  a link  
to the idea of \cite{Abreu2012}
that variations of the adiabaticity, 
and hence of the field storage capacity, in the 
tachocline could explain 
the effect of weak tidal forces on 
{\it long-term} variations of the solar dynamo.

The boundary conditions at the north and south pole 
are $A(0,t)=A(\pi,t)=B(0,t)=B(\pi,t)=0$.

This PDE system is solved by a finite-difference scheme
using the Adams-Bashforth method.
We have validated the numerical method by checking the 
convergence and 
comparing it with some results of \cite{Jennings1991} for the
paradigmatic case with $\alpha(\theta)=\alpha_0 \cos(\theta)$ 
and $\omega(\theta)=\omega_0 \sin(\theta)$. 
Even with such a simple model one can obtain butterfly
diagrams, although one has to be careful with their
interpretation. Details can be found in the
Appendix. 

Throughout the rest of the paper, we employ
a $\theta$-dependence of the $\omega$-effect
in the form
\begin{eqnarray}
\omega(\theta)&=&\omega_0 (1-0.939-0.136 \cos^2(\theta)-0.1457 \cos^4(\theta) )\sin(\theta) \, ,
\end{eqnarray}
as derived from  helioseismological measurements
\citep{Charbonneau1999,Charbonneau2010}.
Note that $\omega(\theta)$, 
which is changing sign at $\theta=55^{\circ}$ and 
$125^{\circ}$, is assumed to be 
constant in time. 
We use a plausible value of 
$\omega_0=10000$
which results from taking the measured 460 nHz frequency at the equator,
an estimated tachocline thickness of 1/10 of its approximate  radius
$R=5 \times 10^8$ m, and 
an assumed value of $\eta=7.16 \times 10^7$ m$^2$/s. This 
somewhat peculiar value, which lies close to the upper margin of the 
commonly used values 10$^6$...10$^8$ m$^2$/s \citep{Charbonneau2010} 
corresponds to a diffusion time $\tau=R^2/\eta=110.7$ years,
which is just a factor 10 times larger than the period of the 
tidal forcing. 

Much less than for $\omega(\theta)$ is known for the 
corresponding distribution of the $\alpha$ effect which we, 
in general, suppose to comprise a non-periodic 
part $\alpha^c$ and a time-periodic part 
$\alpha^p$. 
The non-periodic contribution $\alpha^c$ represents the traditional 
$\alpha$ effect, which is related to the non-mirror symmetric part
of the turbulence. It will be equipped with the typical north-south 
asymmetry and a simple algebraic quenching with the magnetic 
field strength, as it has been utilized in many solar dynamo models. 
The oscillatory contribution $\alpha^p$, however, relies on the 
observation \citep{Weber2015} 
that the TI at low magnetic Prandtl numbers (which applies to the
tachocline) has a
tendency to undergo oscillations of the helicity,
and  that those helicity oscillations
can be resonantly excited by $m=2$ tidal-like perturbations
\citep{Stefani2016}, without (or barely) changing the
energy content of the instability.
The specific forms of both parts of $\alpha$ 
will be discussed further below. At any rate, 
the saturation of the dynamo is exclusively 
accomplished by the
magnetic field dependence, i.e. the quenching of $\alpha$,
while $\omega$ remains unchanged, as stated above.

\section{Synchronizing a pure Tayler-Spruit dynamo model}

In this  section, we illustrate the variety of 
dynamo solutions that arise under
the influence of an $\alpha$-effect that is supposed to 
oscillate with an 11.07 years period and to have a 
specific $B$-dependent amplitude which reflects the 
resonance condition of the periodic tidal trigger with the 
intrinsic oscillation of the TI-related $\alpha$ effect 
\citep{Weber2015,Stefani2016}. 
By virtue of this $B$-dependence of $\alpha$, this model 
can only yield sub-critical dynamo action, 
a fact that will be proven in 
the following. The specific 
effects of combining the periodic $\alpha$-term with a 
more conventional, non-periodic 
$\alpha$-term will be assessed in the
next section.

\subsection{Specifying the $\alpha$-effect}

The time-periodic part
$\alpha^p$ is actually at the
root of our synchronization model. A serious uncertainty
applies to the $\theta$-dependence of this 
term in general, and its equatorial symmetry/asymmetry in 
particular.
A closely related issue is its  ``smoothing''
character, i.e. whether and how
$\alpha^p(\theta,t)$ depends also on $B$ at neighbouring
latitudes and previous times. 

As a first  attempt, we will use an 
$\alpha^p$-dependence on $B$ that is instantaneous 
in $t$ and  local in $\theta$, the latter assumption corresponding 
to a sort of flux-tube approximation. In reality, some 
averaging over time and space, realized by  integral 
kernels, seems more appropriate. Any concretization of
this idea is, however, left  for future work.

As for the latitudinal symmetry property of $\alpha^p$ 
we will start with the plausible assumption 
that it has the same north-south asymmetry as 
is usually assumed for the
non-periodic part. This relies on the observation
of \cite{Ruediger2013} that, under the additional 
influence of a poloidal field, the helicity of the
TI-related $\alpha$-effect is governed by the pseudo-scalar
${\bf B} \cdot  \nabla \times {\bf B}$ (rather than by the pseudo-scalar 
${\bf g} \cdot \nabla \times {\bf \Omega}$, formed with the 
stratification vector 
${\bf g}$ and the global rotation ${\bf \Omega}$).
Although this argument applies, first of all, to the
non-oscillatory part of $\alpha$ for which it 
predicts a positive value in the northern and a negative 
value in the southern hemisphere, we extend here this
equatorial asymmetry also to the oscillatory part. 
That this is in no way self-evident, and should be 
scrutinized in future work, can be inferred from the work
of \cite{Proctor2007} who obtained for his 
fluctuating $\alpha-\Omega$ model an averaged
induction term that is
{\it symmetric} about the equator. 

In contrast to \cite{Ruediger2013}, we 
further assume that $\alpha^p$ is restricted to
the  $\pm 35^{\circ}$ strip 
around the equator, since this is the region with 
positive radial 
shear where the TI may have time to develop, 
not being overrun by the faster magnetorotational instability 
(MRI) that might be dominant 
in the near-pole regions characterized by negative radial shear 
\citep{Kagan2014,Jouve2015}. 
While this restriction  to the $\pm 35^{\circ}$ strip
sounds plausible also with view on the restriction of
sunspots to this area, with regard to the key role
of the $\pm 55^{\circ}$ latitude region for starting the dynamo cycle
\citep{McIntosh2015}, the entire argument might not be 
that convincing.
We will come back to this point in the conclusions.
 
Thus motivated, we start with the following 
parametrization for $\alpha^p(\theta,t)$:
\begin{eqnarray}
\alpha^p(\theta,t)&=&\alpha^p_0 \sin(2 \pi t/11.07) 
 \frac{B^2(\theta,t)}{(1+q^p_{\alpha} 
B^4(\theta,t))} S(\theta) \; \mbox{for $55^{\circ}<\theta<125^{\circ}$} \nonumber \\
&=&0 \; \mbox{elsewhere} \; ,
\end{eqnarray}
where the $B$-dependent term is supposed to have the 
typical resonance-type 
structure $\sim B^2/(1+q^p_{\alpha} B^4)$ 
as already used in the ODE system \citep{Stefani2016}.
Note that the latitudinal dependence 
\begin{eqnarray} 
S(\theta)&=&
{\rm{sgn}}(90^{\circ}-\theta)  \nonumber \\
&\times& \left[ 1-\left( 1+\tanh\left(  \frac{\theta/180^{\circ}-0.5}{0.2}  \right)    \right)
                 \left( 1-\tanh\left(  \frac{\theta/180^{\circ}-0.5}{0.2}   \right)    \right) \right]
\end{eqnarray}
comprises a smoothing term
around the equator 
in order to avoid a numerically inconvenient 
steep jump of $\alpha$ here. 

At any rate, $\alpha^p(\theta,t)$ is not  pre-given but  
co-evolves with the solution of the PDE system. For 
its interpretation we recall the connection to the dimensional
value, $\alpha=\alpha_{\dim} R/\eta$, which leads (with 
$R=5 \times 10^8$ m , $\eta=7.16 \times 10^7$ m$^2$/s) to 
$\alpha_{\rm dim}=\alpha/6.98 $ m/s. That is, all values shown in the 
following figures should be divided by a factor 7 to get the 
physical value $\alpha_{\rm dim}$ in m/s.
Note that, since we have used a comparably high value of $\eta$,
the resulting values of $\alpha_{\rm dim}$ should be considered 
an upper limit and
might in reality be significantly smaller.
The constant term $\alpha^c_0$ is set to a very small, 
but non-zero value of 0.001.

Since for the sub-critical dynamo type to be studied here the
initial conditions play an essential role, we state  them 
explicitly:
\begin{eqnarray}
A(\theta,0)&=&s \sin(\theta) + u \sin(2 \theta)\\
B(\theta,0)&=&-s \sin(2 \theta) - u \sin(\theta) \; .
\end{eqnarray}
Both pre-factors $s$ and $u$, which  denote symmetric and asymmetric 
components for $A$, are usually set to some non-zero value, 
in order not to suppress artificially any relevant modes.

\subsection{The case $\kappa=0$}

\begin{figure}[!ht]
\includegraphics[width=120mm]{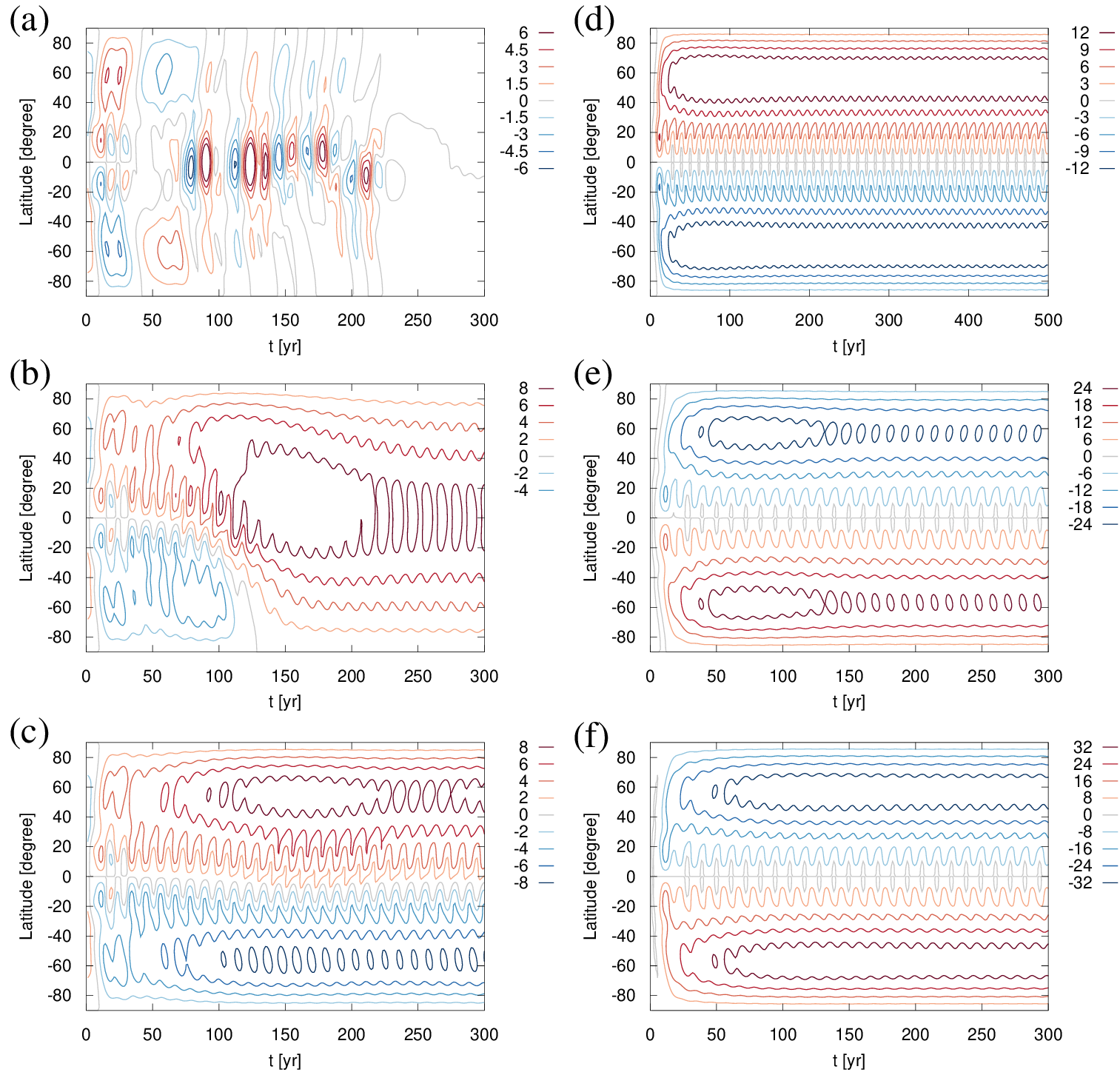}
\caption{Behaviour of $B(\theta,t)$ of 
the synchronized 
Tayler-Spruit dynamo with a nearly pure periodic $\alpha^p$ term.	     
The fixed parameters are $\omega_0=10000$, $\kappa=0$, $\alpha^c_0=0.001$
$q^p_{\alpha}=0.2$,
the initial conditions are $s=3$ and $u=0.001$, 
and the varying parameter is 
$\alpha^p_0=$ 16.1 (a), 16.2 (b), 16.3 (c), 30 (d), 70 (e), 150 (f).
Note that, here and throughout the paper, 
the ordinate axis represents not the co-latitude $\theta$ but the 
normal solar latitude  $90^{\circ}-\theta$.
}
\label{Fig:zusa0_tom_dipol3}
\end{figure}

Figure \ref{Fig:zusa0_tom_dipol3} shows the  
behaviour of $B(\theta,t)$
for the specific parameter choice $\omega_0=10000$, $\kappa=0$,
$q^p_{\alpha}=0.2$, and the initial conditions 
$s=3$ and $u=0.001$, when varying the strength of the
the periodic $\alpha$ term, i.e $\alpha^p_0$ 
between 16.1 (a) and 150 (f).
Evidently, the dynamo starts only for $\alpha^p_0=16.2$ (b), while 
it still dies out for the slightly smaller value $\alpha^p_0=16.1$ (a).

Also interesting is the distinction between a 
quadrupole, pulsating with 11.07 years period,
that arises for $\alpha^p_0=16.2$ (b),
and the pulsating dipole (also with 11.07 years period) 
into which the field evolves for 
$\alpha^p_0=16.3$ (c). This pulsating dipole persists then
also
for the three higher values 
$\alpha^p_0=$ 30 (d), 70 (e), 150 (f).

\begin{figure}[!ht]
\begin{center}
\includegraphics[width=60mm]{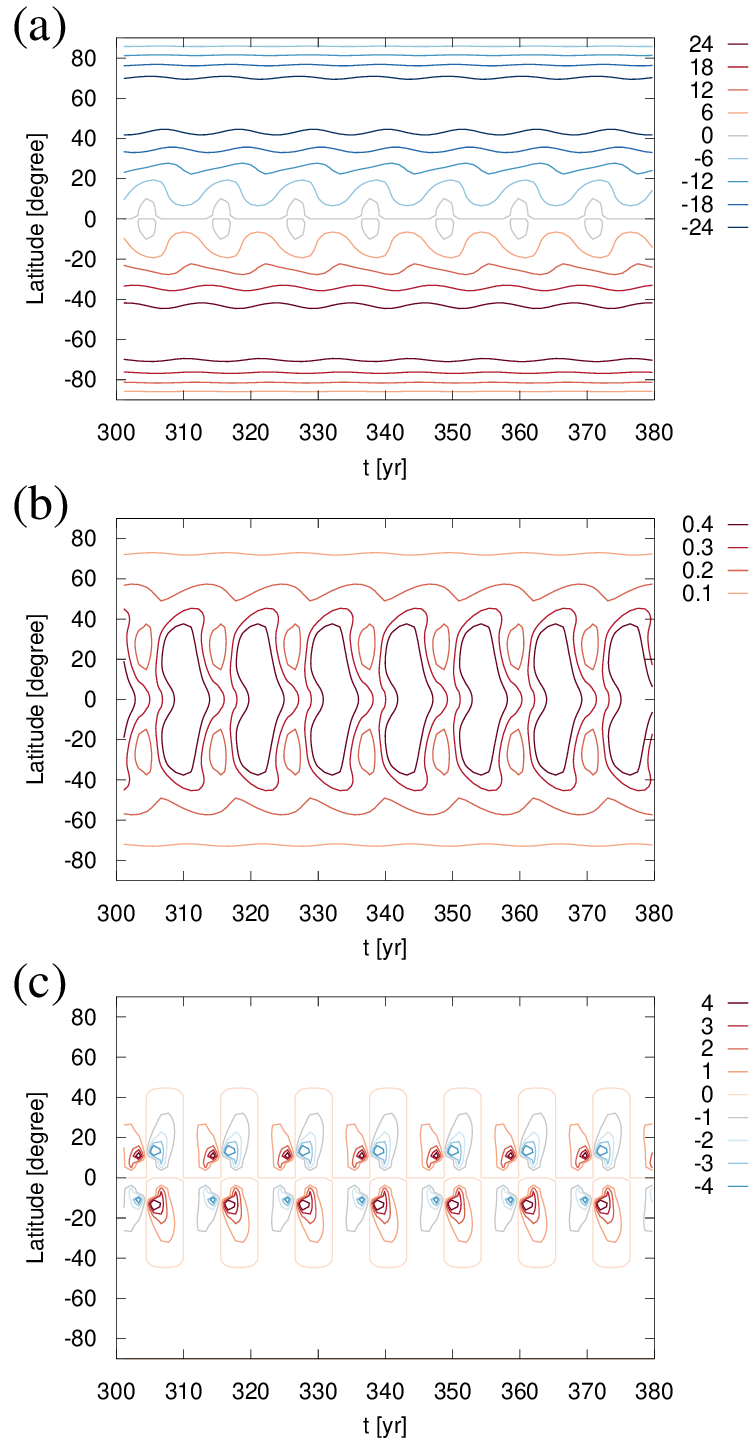}
\caption{Behaviour of $B(\theta,t)$ (a), $A(\theta,t)$ (b), and $\alpha(\theta,t)$ (c).
Parameters as in Figure \ref{Fig:zusa0_tom_dipol3}, but with
$\alpha^p_0=100$.}
\label{Fig:zusa0_faccsi10_dipol3_tom}
\end{center}
\end{figure}

Some detailed features of this dynamo behaviour are illustrated in 
Figure \ref{Fig:zusa0_faccsi10_dipol3_tom} for another value 
$\alpha^p_0=100$ 
(which would lie between panels (e) and (f) of Figure 
\ref{Fig:zusa0_tom_dipol3}).
Complementary to $B$ (a), the poloidal field $A$ (b) shows clearly the 
pulsating dipolar field structure. 
While $\omega$
is kept constant over time (see Equation (3)), 
the behaviour of $\alpha(\theta,t)$  (c)
is more interesting:
Restricted, by construction, to the $\pm 35^{\circ}$ strip around the
equator  (i.e. $55^{\circ}<\theta<125^{\circ}$),
its dependence on $B$ leads to 
typical sign changes in both hemispheres, 
a feature that could possibly be linked to the
reversed current helicity as intermittently 
observed on the sun \citep{Zhang2010}.

\subsection{The case $\kappa \ne 0$}

\begin{figure}[!ht]
\includegraphics[width=120mm]{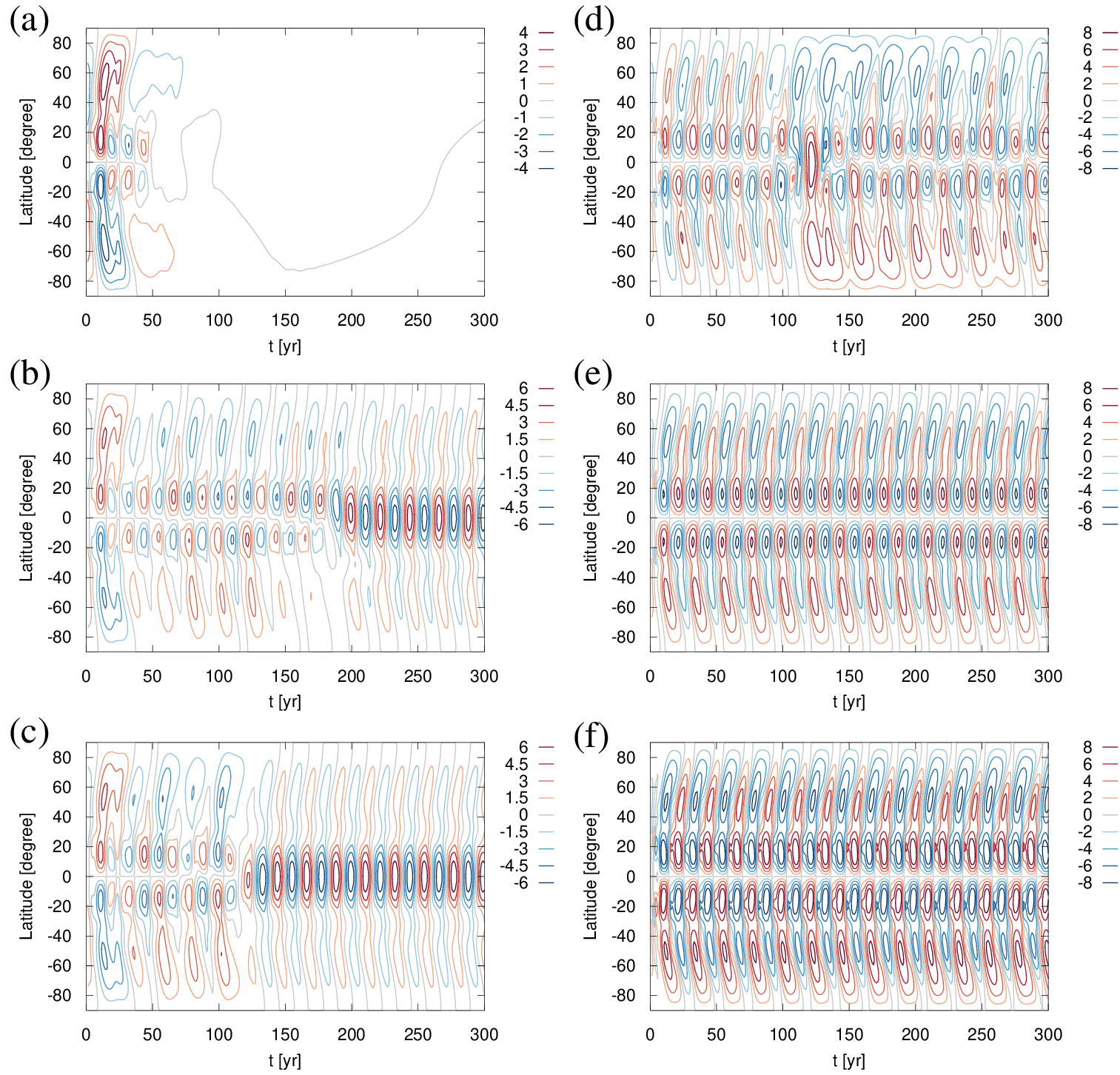}
\caption{Behaviour of $B(\theta,t)$ of 
the synchronized 
Tayler-Spruit dynamo with a periodic $\alpha^p$ term.	     
The fixed parameters are $\omega_0=10000$, $\kappa=1$, 
$q^p_{\alpha}=0.2$,
the initial conditions are $s=3$ and $u=0.001$, 
and the varying parameter is 
$\alpha^p_0=$ 21.2 (a), 21.5 (b), 23 (c), 50 (d), 70 (e), 150 (f).
}
\label{Fig:zusa1_tom_dipol3}
\end{figure}

Up to this point, the final state of the dynamo was, 
somewhat disappointing,  
either a pulsating quadrupole or a pulsating dipole. In the 
following we will also find oscillatory dipoles when 
going over to $\kappa \ne 0$, i.e. when allowing for some
magnetic field loss due to rising flux tubes. 
The results are illustrated for the specific choice 
$\kappa=1$. With all remaining parameters unchanged
(i.e. $\omega_0=10000$, $q^p_{\alpha}=0.2$, $s=3$ and $u=0.001$),
Figure \ref{Fig:zusa1_tom_dipol3} shows the behaviour of
$B(\theta,t)$ when varying the amplitude of $\alpha^p_0$ now
 between 21.2 (a) and 150 (f).
Evidently, since the additional 
field losses have to be compensated, 
the dynamo starts now only for $\alpha^p_0=21.5$ (b), 
while dying out for the slightly smaller value $\alpha^p_0=21.2$ (a).

Whereas for $\alpha^p_0=21.5$ (b) and $\alpha^p_0=23$ (c) the initially
prescribed dipole finally gives way to a quadrupole oscillating with
22.14 years period, for $\alpha^p_0=50$ (d) it recovers after a short
excursion (between 110...130 years) to a 
hemispherical and quadrupolar mode. 
While such spontaneous dipole-quadrupole transitions 
are found here only in certain parameter regions, 
we will later see that they 
can be easily triggered by changing such parameters as the 
amplitude of  $\alpha^p$ 
or the loss parameter $\kappa$. 
For $\alpha^p_0=70$ (e) and
$\alpha^p_0=150$ (f) we obtain very regular 
dipole oscillations, although 
in either case with a clear Gnevyshev-Ohl tendency.

\begin{figure}[!ht]
\begin{center}
\includegraphics[width=60mm]{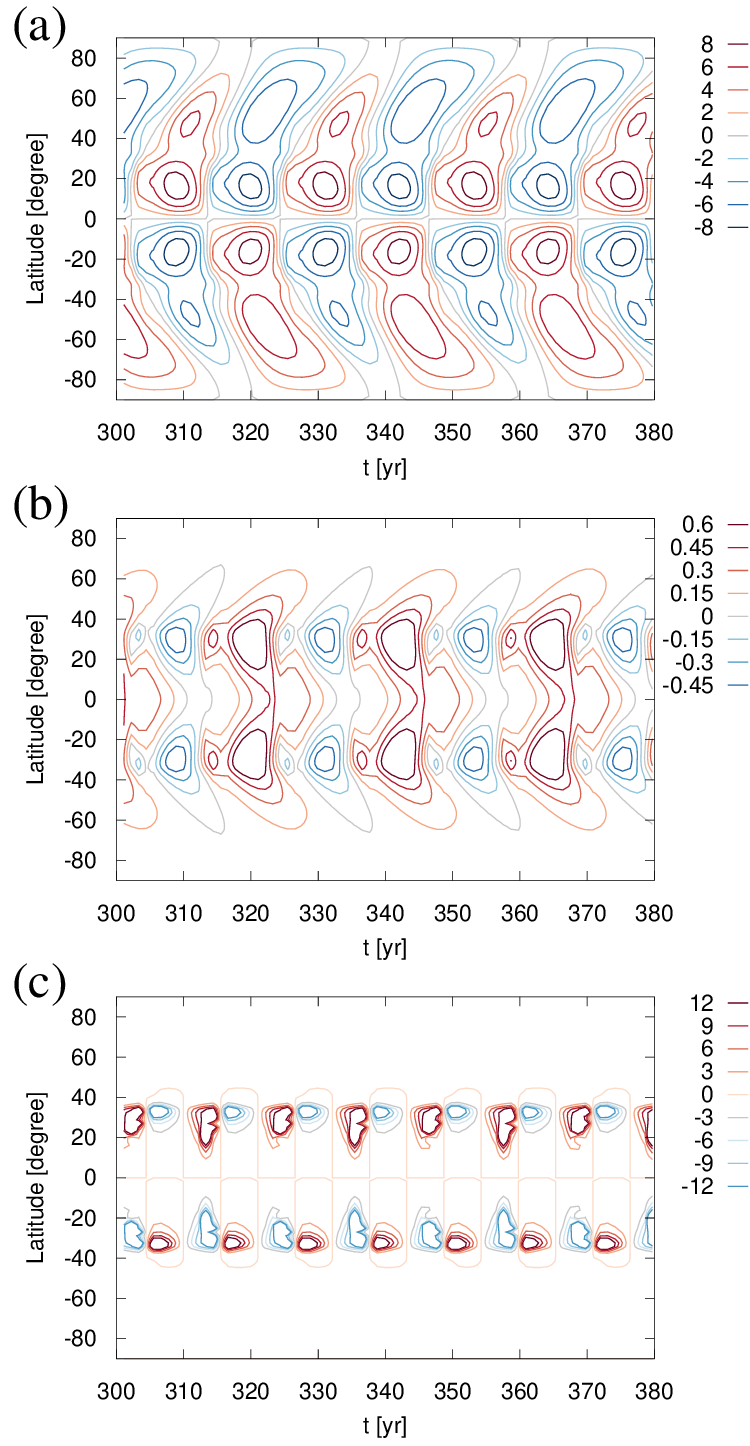}
\caption{Behaviour of $B(\theta,t)$ (a), $A(\theta,t)$ (b), and $\alpha(\theta,t)$ (c).
Parameters as in Figure \ref{Fig:zusa1_tom_dipol3}, but with
$\alpha^p_0=100$.
}
\label{Fig:zusa1_faccsi10_dipol3_tom}
\end{center}
\end{figure}

Again, we illustrate in Figure \ref{Fig:zusa1_faccsi10_dipol3_tom}
the detailed behaviour for the particular value
$\alpha^p_0=100$, which lies between panels 
(e) and (f) of Figure \ref{Fig:zusa1_tom_dipol3}.
Actually, the results exhibit some interesting features which are not untypical
for the sun. First, (a) shows for high latitudes 
the typical ``rush to the poles'', while
for low latitudes we see a sort of butterfly slightly tending 
equator-ward. Admittedly, the shape of this butterfly is not
convincing yet, and it remains to be seen whether this shape 
can be improved in higher-dimensional simulations, 
including also the meridional circulation.

Second, the Gnevyshev-Ohl tendency
becomes clearly visible with the ``blue field'' in the northern hemisphere
being stronger than the ``red field'' (and vice versa in the southern 
hemisphere). Closely related to that feature, 
the $\alpha$ values in (c) show also some 
symmetry breaking between positive and negative values.

\subsection{The subcritical character of the Tayler-Spruit dynamo}

\begin{figure}[!ht]
\includegraphics[width=120mm]{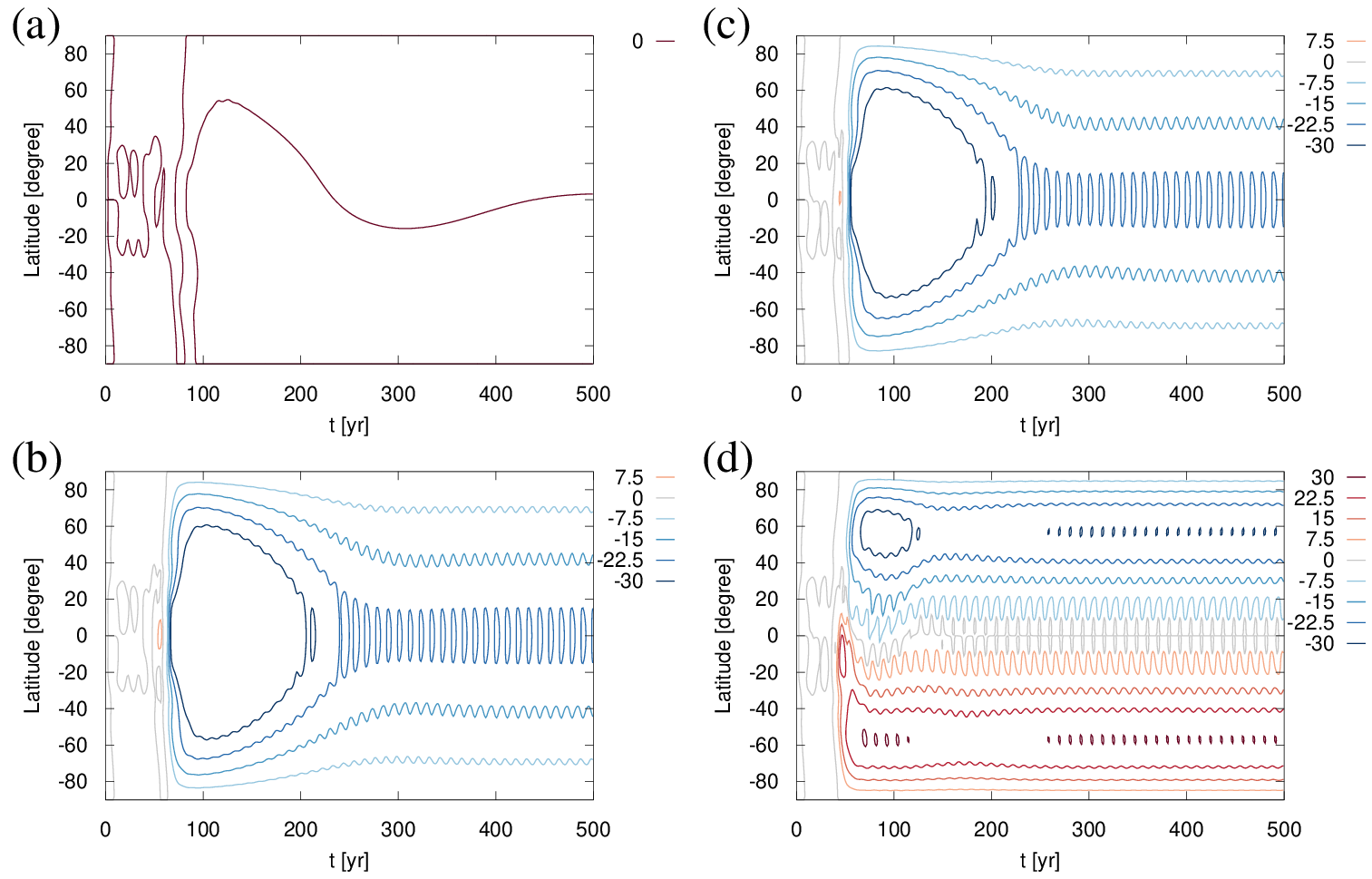}
\caption{Behaviour of $B(\theta,t)$ of the synchronized 
$\alpha-\Omega$ model 
with the fixed values $\Omega_0=10000$, $\alpha^p_0=100$, 
$\kappa=0$, $q^p_{\alpha}=0.2$, $u=0.001$  and the 
variable initial conditions 
$s=0.707$ (a), 0.708 (b), 0.729 (c) and 0.73 (d).}
\label{Fig:zusa0_faccsi10_subcritical}
\end{figure}

Now we address the subcritical nature of the dynamo which is, 
in terms of a high sensitivity on the
initial conditions, illustrated in Figure 
\ref{Fig:zusa0_faccsi10_subcritical}.
We choose again $\omega_0=10000$, $\kappa=0$,
$q^p_{\alpha}=0.2$, $u=0.001$, but vary now the initial value
of the dipole strength $s$ in a narrow interval between 0.707 and 0.73.
The value of $\alpha^p_0=100$ is chosen 
to lie between 70  (cp. Figure \ref{Fig:zusa0_tom_dipol3}(e)) 
and 150 (Figure \ref{Fig:zusa0_tom_dipol3}(f)).
Remarkably, the dynamo starts only  
when $s \ge 0.708$ (b), while
the slightly weaker initial perturbation $s=0.707$ (a) dies away
at  large times. Further to this, between $s=0.729$ (c) and 
$s=0.73$ (d) the dynamo field changes from a pulsating quadrupole to
a pulsating dipole.

\begin{figure}[!ht]
\begin{center}
\includegraphics[width=100mm]{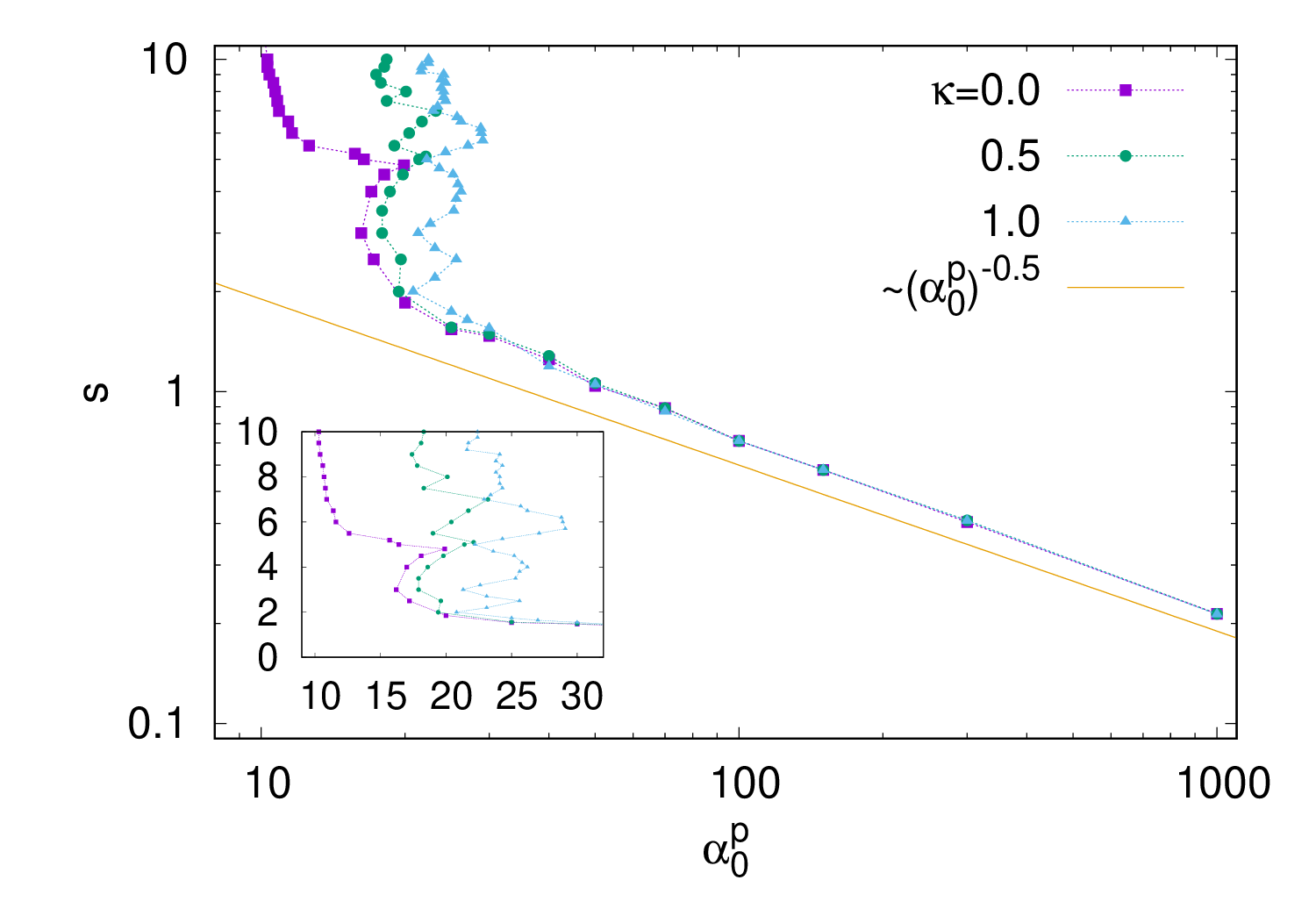}
\caption{Stability boundaries in the $\alpha^p_0-s$ plane, for 
the three values 
$\kappa=0$, 0.5 and 1 and $q^p_{\alpha}=0.2$, $u=0.001$. 
Note the left ``rugged'' 
boundary at low values of  $\alpha^p_0$. For large 
values of  $\alpha^p_0$, the boundary 
converges to $s \sim (\alpha^p_0)^{-0.5}$. 
}
\label{Fig:grenze_zusammen}
\end{center}
\end{figure}

The subcritical behaviour is summarized  in Figure 
\ref{Fig:grenze_zusammen} which shows the dynamo threshold
in the $\alpha^p_0-s$ plane, for the three specific loss 
parameters $\kappa=0$, 0.5 and 1. Each of the points in this
graphic has been determined by evaluating the dynamo/non-dynamo
behaviour at a few points in its vicinity. For large 
values of $\alpha^p_0$ we obtain the typical subcritical 
$s \sim (\alpha^p_0)^{-0.5}$ behaviour,
which means that the necessary initial condition can be lowered 
(with the square-root) when the dynamo strength is increased.
Also typical for a subcritical bifurcation is the
``rugged'' left boundary, which is reminiscent of a similar
fractal shape found for pipe flows \citep{Eckhardt2008}.
We only mention here that a similar subcritical 
behaviour can also be obtained, with less numerical effort, 
for the ODE case.

\section{Synchronizing a hybrid dynamo}

Having verified the subcritical nature of the pure
Tayler-Spruit model, we will now reinstate the effect
of the more traditional part of $\alpha$ which we parametrize, 
for the sake of convenience, as 
\begin{eqnarray}
\alpha^c(\theta,t)&=&\alpha^c_0(1+\xi(t))  \sin(2 \theta) 
\frac{1}{(1+q^c_{\alpha} B^2(\theta,t))} \; ,
\end{eqnarray}
where $\alpha^c_0$ is a constant and $\xi(t)$ 
denotes a noise term to be specified further below.
The factor $\sin(2 \theta)$ ensures the typical north-south
asymmetry as it is often assumed for conventional 
$\alpha-\Omega$ dynamos. Interestingly, the same 
symmetry argument would also apply to a TI-related,
non-oscillatory $\alpha$ term under the influence 
of an additional poloidal field \citep{Ruediger2013,Ruediger2018}. 
Therefore, any such non-oscillatory contribution 
of the TI-related $\alpha$ effect could 
be consistently absorbed into Equation (8).

\subsection{Noise-free case}

\begin{figure}[!ht]
\includegraphics[width=120mm]{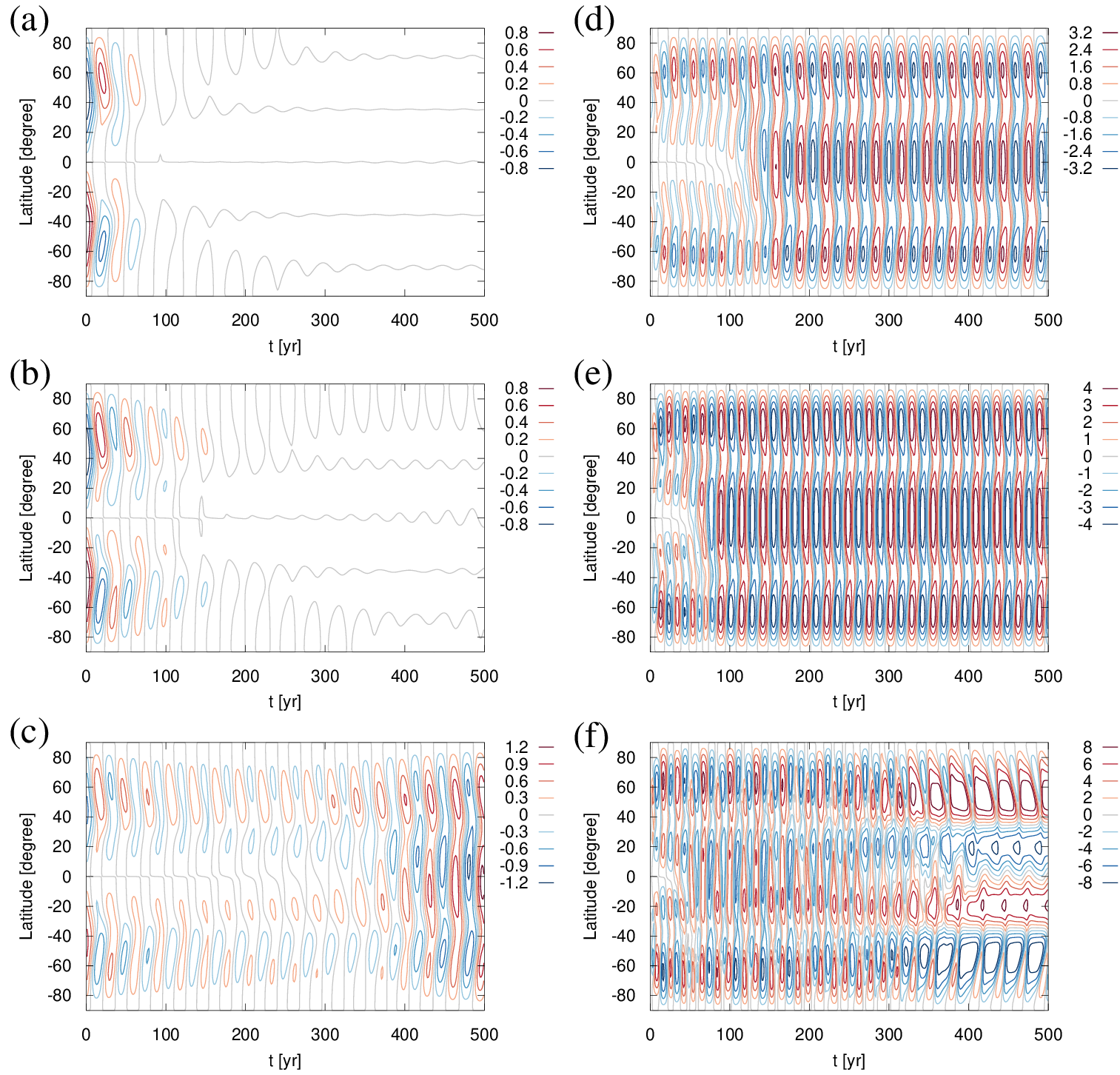}
\caption{Behaviour of $B(\theta,t)$ of 
the traditional $\alpha-\Omega$ dynamo without periodic term, 
i.e. with $\alpha^p_0=0$. The fixed parameters are $\omega_0=10000$, 
$\kappa=0.5$, 
$q^c_{\alpha}=0.8$, $q^p_{\alpha}=0.2$,
the initial conditions are $s=1$ and $u=0.001$, 
and the varying parameter is 
$\alpha^c_0=$ 0.6 (a), 0.8 (b), 1 (c), 4 (d), 10 (e), 40 (f).
}
\label{Fig:zusa0_facco_facsi0_tom_dipol1}
\end{figure}

Let us start with the noise-free case, i.e. $\xi(t)=0$,
for which we consider first a purely 
conventional $\alpha-\Omega$ dynamo,
by skipping the periodic part completely, i.e. by choosing 
$\alpha^p_0=0$.
Figure \ref{Fig:zusa0_facco_facsi0_tom_dipol1} shows the time 
evolution for increasing 
intensity of the constant part, i.e. 
$\alpha^c_0=$ 0.6 (a), 0.8 (b), 1 (c), 4 (d), 10 (e) and
40 (f). 
The other parameters are $\omega_0=10000$, $q^c_{\alpha}=0.8$, 
$\kappa=0.5$. 
While the field clearly dies out for $\alpha^c_0=0.6$ (a), for 
$\alpha^c_0=0.8$ it seems 
to recover very slowly, and for $\alpha^c_0=1$, we get a
clear dynamo with an oscillatory quadrupole which also prevails
for $\alpha^c_0=4$ (d) and $\alpha^c_0=10$ (e). 
At $\alpha^c_0=40$ (e), the dynamo field undergoes several changes and
ends up in a dipole field pulsating with a period 
of approximately 27 years. Note that we have here extended 
the time period to 
500 years in order to show all relevant transitions 
which are partly very slow.

\begin{figure}[!ht]
\begin{center}
\includegraphics[width=80mm]{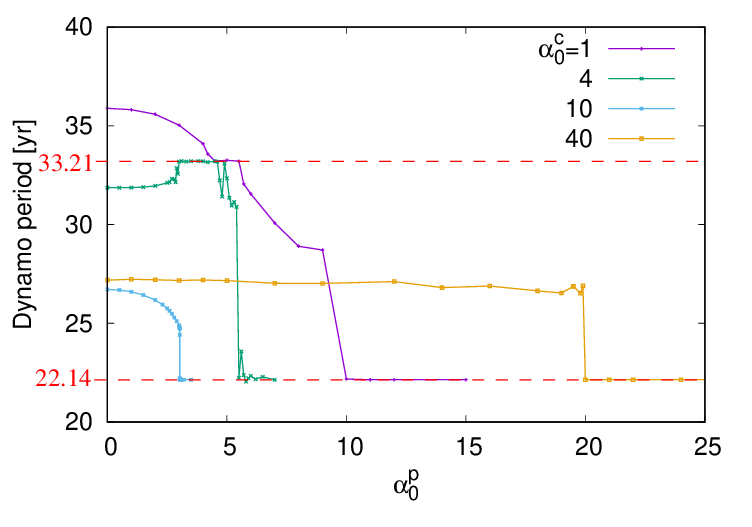}
\caption{Resonance with the external frequency when increasing
$\alpha^p_0$, for four different values of $\alpha^c_0=$ 1, 4, 10, 40,
whose $\alpha^p_0=0$ limit
corresponds to panels (c), (d), (e) and (f) 
of Figure \ref{Fig:zusa0_facco_facsi0_tom_dipol1}, respectively.
}
\label{Fig:resonanz_alle}
\end{center}
\end{figure}

What happens now if we complement this standard 
$\alpha-\Omega$ dynamo  with the periodic $\alpha$ term?
For the four specific choices $\alpha^c_0=$ 1, 4, 10, 40 
(cp. Figure \ref{Fig:zusa0_facco_facsi0_tom_dipol1} (c-f)),
we show in Figure \ref{Fig:resonanz_alle} 
the resulting dynamo period when cranking up the
value of $\alpha^p_0$. For each considered 
value of $\alpha^c_0$, we ultimately 
obtain  a clear synchronization
to a 22.14 years period when the value of $\alpha^p_0$ reaches a 
certain critical value.
In cases that the original period is higher ($\alpha^c_0=$ 1 and 4), 
we also observe an intermediate 2:3 synchronization to a 
33.21 years period. 
Remarkably, the value of $\alpha^p_0$, where the final 
synchronization to 22.14 years is accomplished, 
can be significantly smaller than  the typical 
$\alpha^p_0$ needed for the
pure Tayler-Spruit dynamo to start (cp. 
Figure \ref{Fig:grenze_zusammen}).

\begin{figure}[!ht]
\includegraphics[width=120mm]{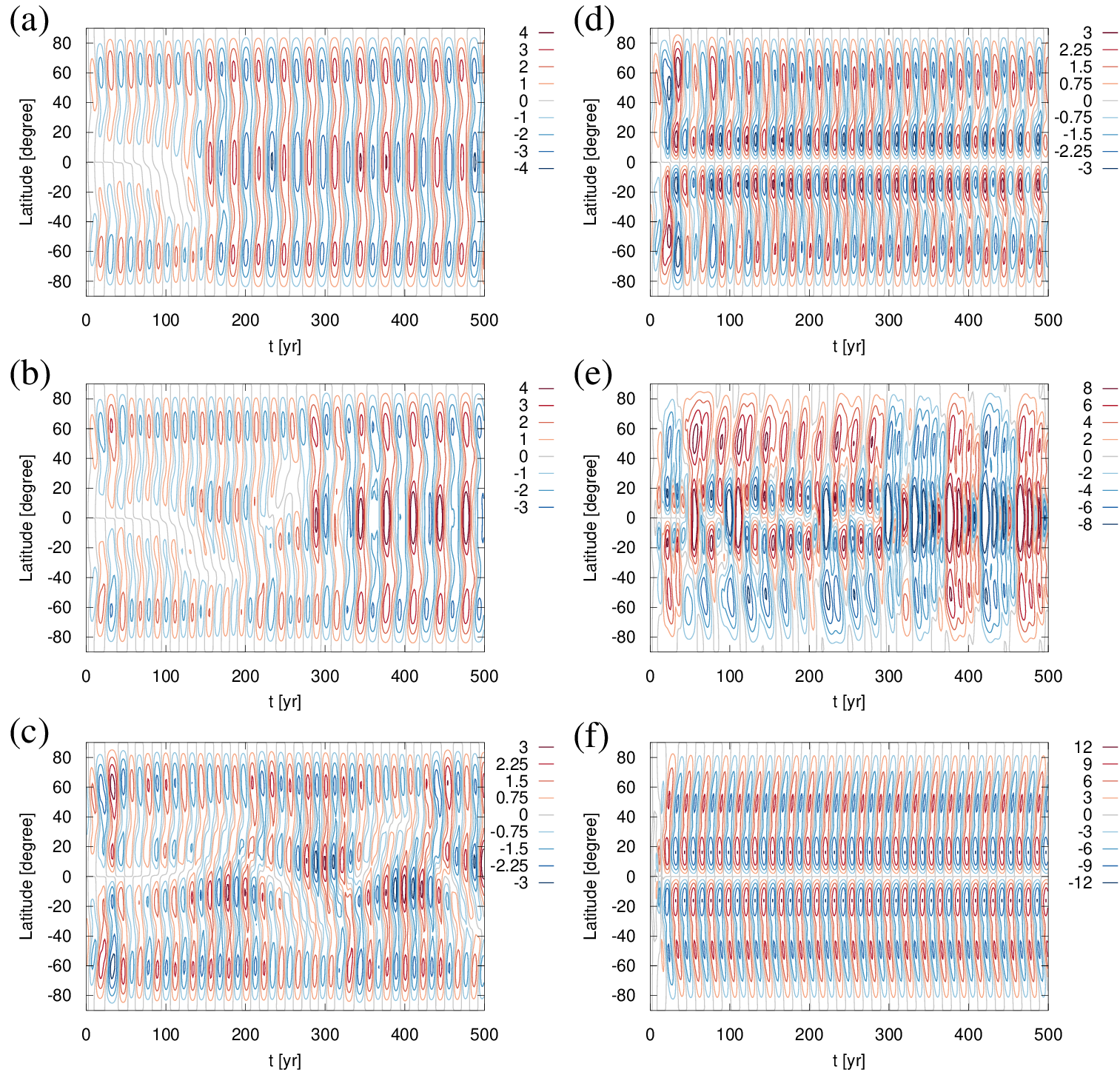}
\caption{Behaviour of $B(\theta,t)$ of 
the traditional $\alpha-\Omega$ combined with increasing $\alpha^p_0$. 
The fixed parameters are $\omega_0=10000$, $\kappa=0.5$, 
$q^c_{\alpha}=0.8$, $q^p_{\alpha}=0.2$, $\alpha^c_0=4$,
the initial conditions are $s=1$ and $u=0.001$, 
and the varying parameter is 
$\alpha^p_0=$ 1 (a), 4 (b), 6 (c), 10 (d), 50 (e), 150 (f).
}
\label{Fig:zusa0_facco04_facsi_tom_dipol1}
\end{figure}

For the specific value $\alpha^c_0=4$ (cp. the green line in Figure
\ref{Fig:resonanz_alle}), Figure \ref{Fig:zusa0_facco04_facsi_tom_dipol1} 
illustrates the complexities of this
synchronization. While for the low value $\alpha^p_0=$ 1 (a) we obtain the
nearly unperturbed oscillatory quadrupole, $\alpha^p_0=$ 4 (b) yields now
the intermediate 2:3 synchronization into a fluctuating 
quadrupole.
Shortly after leaving this 2:3 synchronization regime,
$\alpha^p_0=$ 6 (c) provides a sort of hemispherical field with
22.14 years period, whose
dominating hemisphere is, however, changing with an 
approximately 200 years 
periodicity.
Increasing  $\alpha^p_0$ further to 10 (d), we observe 
a dipole oscillating with
a strong Gnevyshev-Ohl tendency. $\alpha^p_0=50$ produces a wild transition
between oscillatory dipoles and pulsating quadrupoles  at later times.
Very regular dipole oscillations appear then at $\alpha^p_0=150$.
This way, we obtain a transition from
the conventional $\alpha-\Omega$ dynamo via
a hybrid dynamo to a (nearly) pure Tayler-Spruit dynamo,
and synchronization starts at a certain 
fraction of the oscillatory part of $\alpha$.

\begin{figure}[!ht]
\begin{center}
\includegraphics[width=60mm]{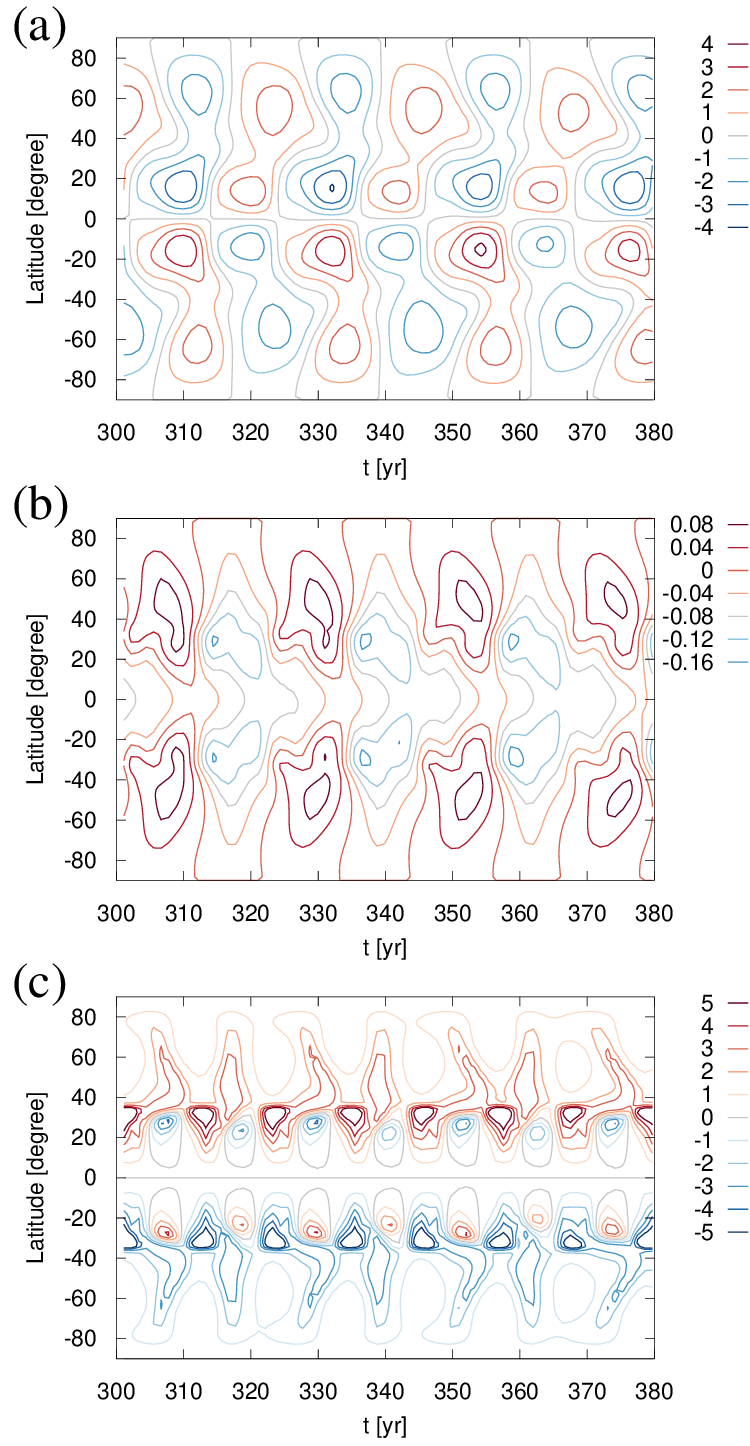}
\caption{Behaviour of $B(\theta,t)$ (a), $A(\theta,t)$ (b), and $\alpha(\theta,t)$ (c).
Parameters as in Figure \ref{Fig:zusa0_facco04_facsi_tom_dipol1}, but with
$\alpha^p_0=12$.
}
\label{Fig:zusa05_facco04_faccsi1k2_dipol1_tom}
\end{center}
\end{figure}

More details of this hybrid dynamo behaviour can be seen
in Figure \ref{Fig:zusa05_facco04_faccsi1k2_dipol1_tom}, documenting the
special case $\alpha^c_0=4$ and  $\alpha^p_0=12$
(similar to Figure \ref{Fig:zusa0_facco04_facsi_tom_dipol1}d).
Here the direction of the butterfly diagram
for low latitudes is not very well expressed.
Quite interesting is the $\alpha$ effect of panel (c)
which shows now, not surprisingly due to the 
presence of $\alpha^c$, a preponderance of positive values in the
northern, and negative values in the southern hemisphere.
The remaining oscillatory part, which has a reasonable 
amplitude 
of approximately 0.5 m/s (recall the necessary division by 7
to get the physical values), is sufficient to synchronize the 
entire dynamo.

\begin{figure}[!ht]
\includegraphics[width=120mm]{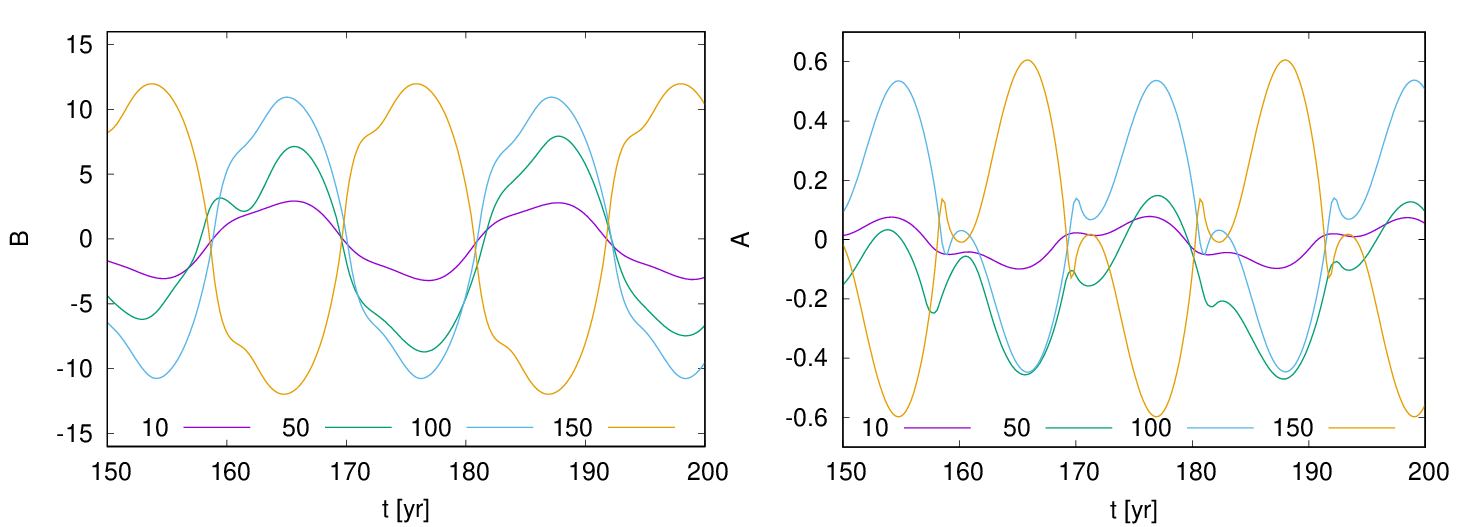}
\caption{Detail of Figure \ref{Fig:zusa05_facco04_faccsi1k2_dipol1_tom} 
for $B(\theta=72^{\circ},t)$ (left) 
and $A(\theta=72^{\circ},t)$ (right) for various values
of $\alpha^p_0$. }
\label{Fig:b40a40}
\end{figure}

Another interesting aspect becomes visible in 
Figure \ref{Fig:zusa05_facco04_faccsi1k2_dipol1_tom}(a,b), and
is quantified in detail in 
Figure \ref{Fig:b40a40} which shows $B(\theta=72^{\circ},t)$
and $A(\theta=72^{\circ},t)$. It refers to the occurrence 
of a double peak of the field amplitude, which is even clearer expressed 
in the poloidal field $A$ than in the toroidal field $B$. This 
double peak is a quite typical feature of the solar dynamo 
and has been discussed, e.g.,  in \cite{Karak2018}.
It might also be worthwhile to 
check the relation of this double peak to the so-called 
''mid-term''periodicities (between 0.5 and 4 years) 
of the solar  activity, as found and discussed by several authors
\citep{Obridko2007,Valdes2008,McIntosh2015,Bazilevskaya2016}.

\subsection{The role of noise}

\begin{figure}[!ht]
\includegraphics[width=120mm]{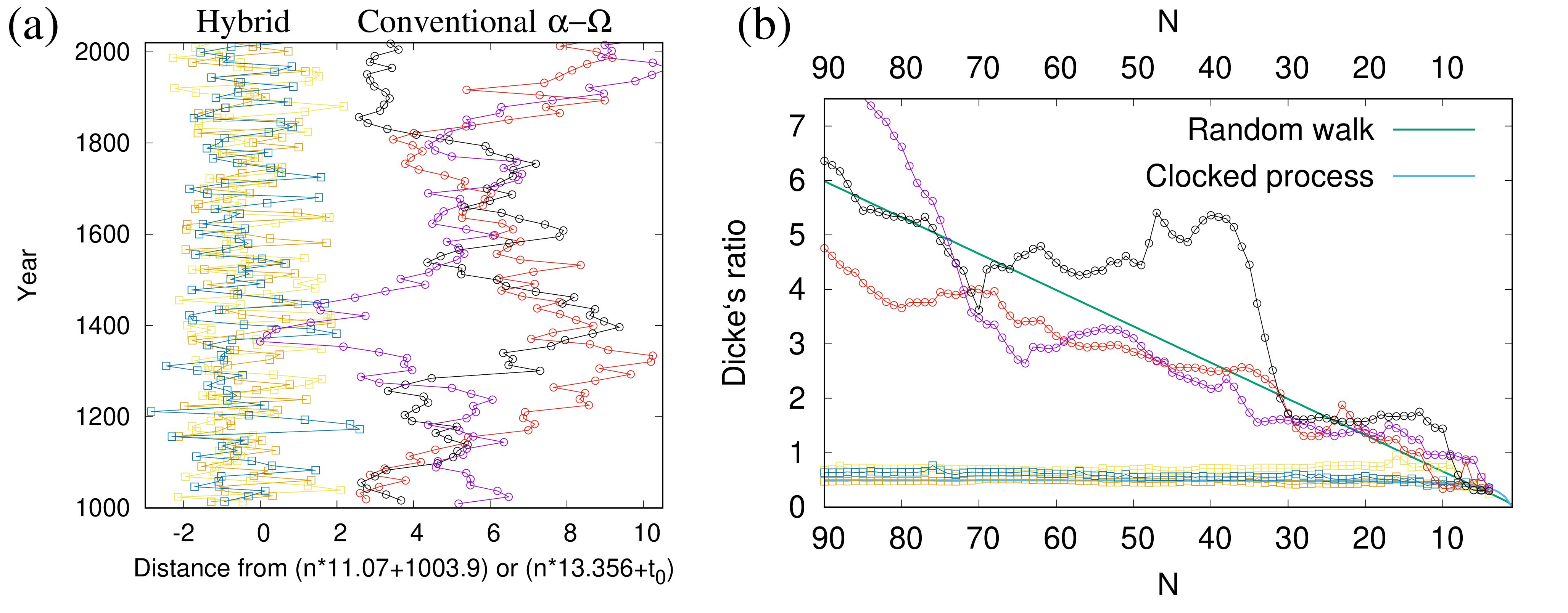}
\caption{The role of noise for the conventional $\alpha-\Omega$ 
and the hybrid  dynamo model. (a) Deviations of 
the time series from linear functions of the cycle number
for three noise realizations in either case. For the conventional 
$\alpha-\Omega$ the three time series (with circles, on the right side) 
undergo long-term excursions, while the time-series for the hybrid dynamo
(with squares, on the left side) remain much closer to the  linear trend. 
(b) Dicke's ratio  in dependence on the
number $N$ of cycles taken into account, 
for a random walk process (green line, converging towards $N/15$), a
clocked process (blue line, converging towards 0.5), and 
the two triples of time series as shown in (a). }
\label{Fig:noise}
\end{figure}

Having seen that a conventional $\alpha-\Omega$ dynamo
with an intrinsic frequency can be synchronized by adding a
periodic $\alpha$ term, we ask now about the specific influence
of noise on the behaviour of these two types of models.
In either case, we augment the non-periodic part of $\alpha$ 
by a noise term $\xi(t)$ defined by the correlator
$\langle \xi(t) \xi(t+t_1) \rangle = D^2 (1-|t_1|/t_{\rm corr})
\Theta(1-|t_1|/t_{\rm corr})$, which is numerically realized by
random numbers with variance $D^2$ which are held constant 
over a correlation time $t_{\rm corr}$.
In the following, we will choose, somewhat arbitrarily, 
$t_{\rm corr}=0.55$ years, which is at any rate significantly 
shorter than the solar cycle.
We start with a pure $\alpha-\Omega$ model with $\alpha^p_0=0$, 
$\alpha^c_0=10$, 
$\omega_0=10000$, $q^c_{\alpha}=0.8$, $q^p_{\alpha}=0.2$, 
$\kappa=0.5$, 
which corresponds to the leftmost point of the blue curve 
in  Figure \ref{Fig:resonanz_alle}.
For $D=0.3$, the rightmost curves (marked by circles) of
Figure \ref{Fig:noise}(a) illustrate three specific
noise realizations, which all exhibit long-term, large-amplitude 
excursions around their linear trends (note that we have used, for the
sake of easy comparison, the same scales as in Figure 
\ref{Fig:vergleich}).
Dicke's ratio for these three curves is shown then, using the same colours,
in Figure \ref{Fig:noise}(b). Despite large deviations of the 
individual curves, we observe a clear resemblance to the 
$\sim N/15$ dependence as typical for a random walk process.

Things are different, though, for the hybrid dynamo. In addition 
to the parameters indicated above, we choose now $\alpha^p_0=5$, which
lies well in the synchronized part of the blue 
curve of  Figure \ref{Fig:resonanz_alle}. The resulting 
three leftmost time-series (marked by squares) in Figure \ref{Fig:noise}(a) 
now remain much closer to the linear trend, without undergoing long-term 
excursions. It is evident, however, that the noise 
alleviates any {\it local} clocking with the periodic forcing, while
the {\it global} clocking is well maintained. Unsurprisingly, 
Dicke's ratio for these time series in Figure \ref{Fig:noise}(b)
is quite close to the ideal curve for a clocked process.
This clear difference between a random walk process and a clocked process,
as evidenced in our two numerical models, makes it indeed worthwhile
to validate or improve Schove's data 
on which the curves in Figures \ref{Fig:vergleich} and \ref{Fig:dicke}
were based on.

\section{Modeling grand minima}

In contrast to the idea of a {\it hard synchronization} of the
basic Hale cycle with planetary tidal 
forces, as pursued in this paper,  much 
more interest is commonly devoted to the possibility of a 
{\it soft modulation} of the solar activity, with 
particular focus on the  Gleissberg, 
Suess-de Vries, Hallstadt, and Eddy cycles
\citep{Jose1965,Charvatova1997,Abreu2012,Wolf2010,Scafetta2010,Scafetta2014,McCracken2014,Cionco2015,Scafetta2016}.
While far from being settled 
(see, e.g.,  \cite{Cameron2013} for a critical
assessment), any such planetary influence could
have enormous consequences for the predictability not only 
of the solar dynamo but, possibly,
of the terrestrial climate, too
\citep{Hoyt1997,Gray2010,Solanki2013,Scafetta2013,Ruzmaikin2015,Soon2014}.
It is, therefore, worthwhile to figure out whether our model
can explain modulations of the solar cycle, 
including extreme cases such as the Maunder and 
other grand minima.

We had already seen above (Figure \ref{Fig:zusa1_tom_dipol3}d and 
Figure \ref{Fig:zusa0_facco04_facsi_tom_dipol1}e) that for 
some parameter choices transitions between dipoles and quadrupoles 
can even occur spontaneously, which indicates a high
sensitivity of the corresponding dynamo with respect to
minor parameter variations.
Based on this observation, we study here the transition
between the two field topologies when allowing the ratio
of $\alpha^p$ to $\alpha^c$ to vary with a long period, for which 
we take here 550 years just for the sake of concreteness
(at a comparably 506 years period, \cite{Abreu2012} found a 
peak both in the solar modulation potential and the annually 
averaged planetary torque modulus).
 
\begin{figure}[!ht]
\includegraphics[width=120mm]{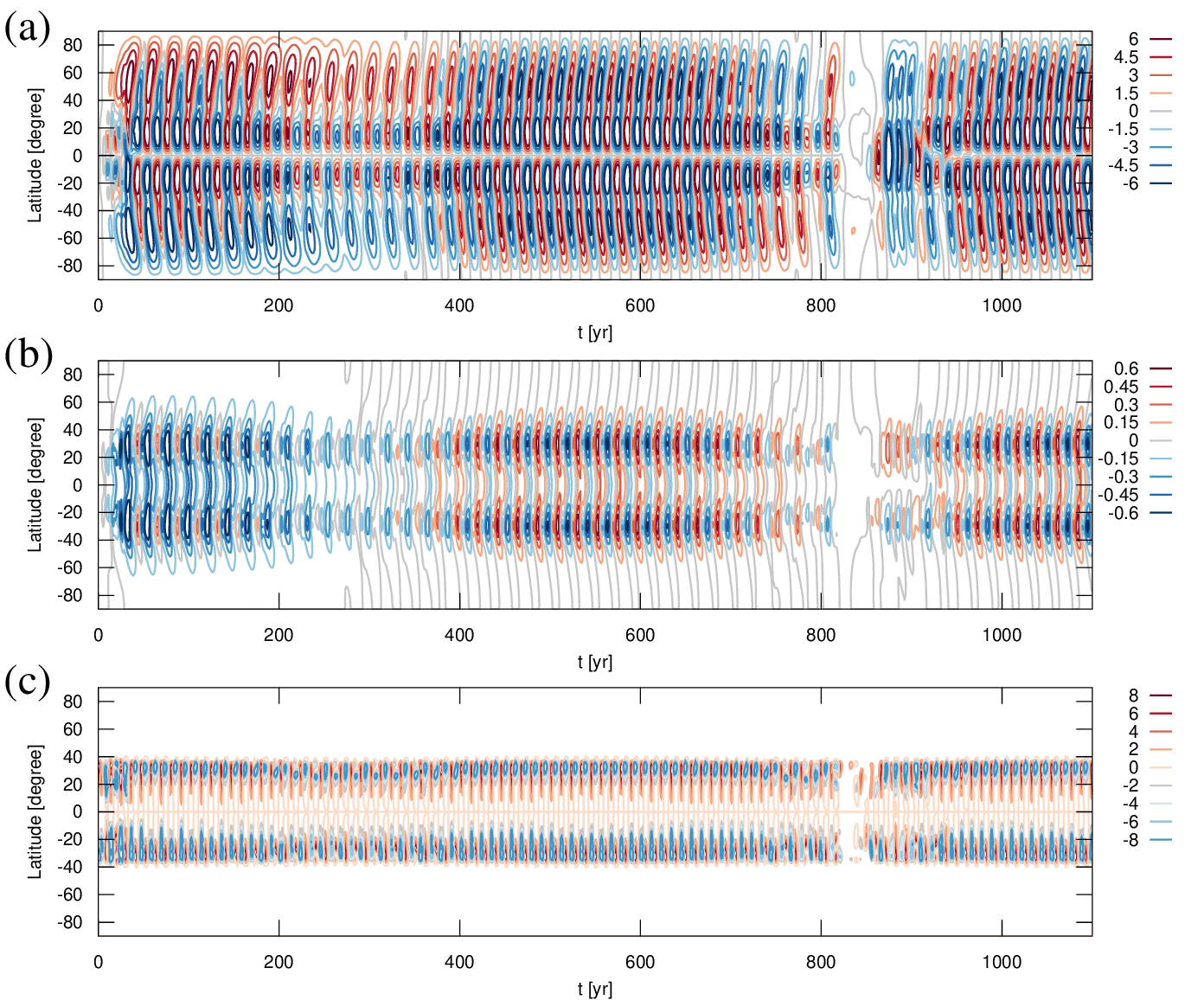}
\caption{Behaviour of $B(\theta,t)$ (a), $A(\theta,t)$ (b), and 
$\alpha(\theta,t)$ (c) showing transitions between  
dipole and quadrupole fields
when varying $\alpha^p_0$ according to $\alpha^p_0(t)=90(1-0.7 \sin^2(2 \pi t/1100))$. The 
fixed parameters are $\Omega_0=10000$, $\alpha^c_0=1$, $\kappa=1$
$q^p_{\alpha}=0.2$, $q^c_{\alpha}=0.8$.}
\label{Fig:maunder1}
\end{figure}

Using the fixed parameters $\omega_0=10000$, $\alpha^c_0=1$, $\kappa=1$
$q^p_{\alpha}=0.2$, $q^c_{\alpha}=0.8$,
we consider now $\alpha^p_0$ in Equation (4) as time-dependent and 
vary its value between 27 and
90 according to $\alpha^p_0(t)=90(1-0.7 \sin^2(2 \pi t/1100))$.
This function 
has maxima at $t=$ 0, 550 and 1100, and minima at
$t=$ 225 and 775.
Figure \ref{Fig:maunder1} shows the results: at the first minimum
of $\alpha^p_0$,
around t=225, the dipolar field is just weakened and does
not undergo a transition to a quadrupole, while exactly this 
happens at the second minimum,
after t=775, where  the dipole shortly vanishes and gives way to 
a quadrupole field before coming back again around $t=900$. 
This difference in behaviour at the first and second minimum
of $\alpha^p_0$ indicates a high sensitivity of 
these transitions.
Note that in particular the transition between quadrupole and dipole looks
similar to that after the Maunder minimum (\cite{Arlt2009,Moss2017}).

Most important here is the phase memory during all 
these transitions. This feature brings us back 
to the amazing persistence of the solar cycle, 
even during the Maunder minimum,
as it was demonstrated in Figure \ref{Fig:vergleich}.

\begin{figure}[!ht]
\includegraphics[width=120mm]{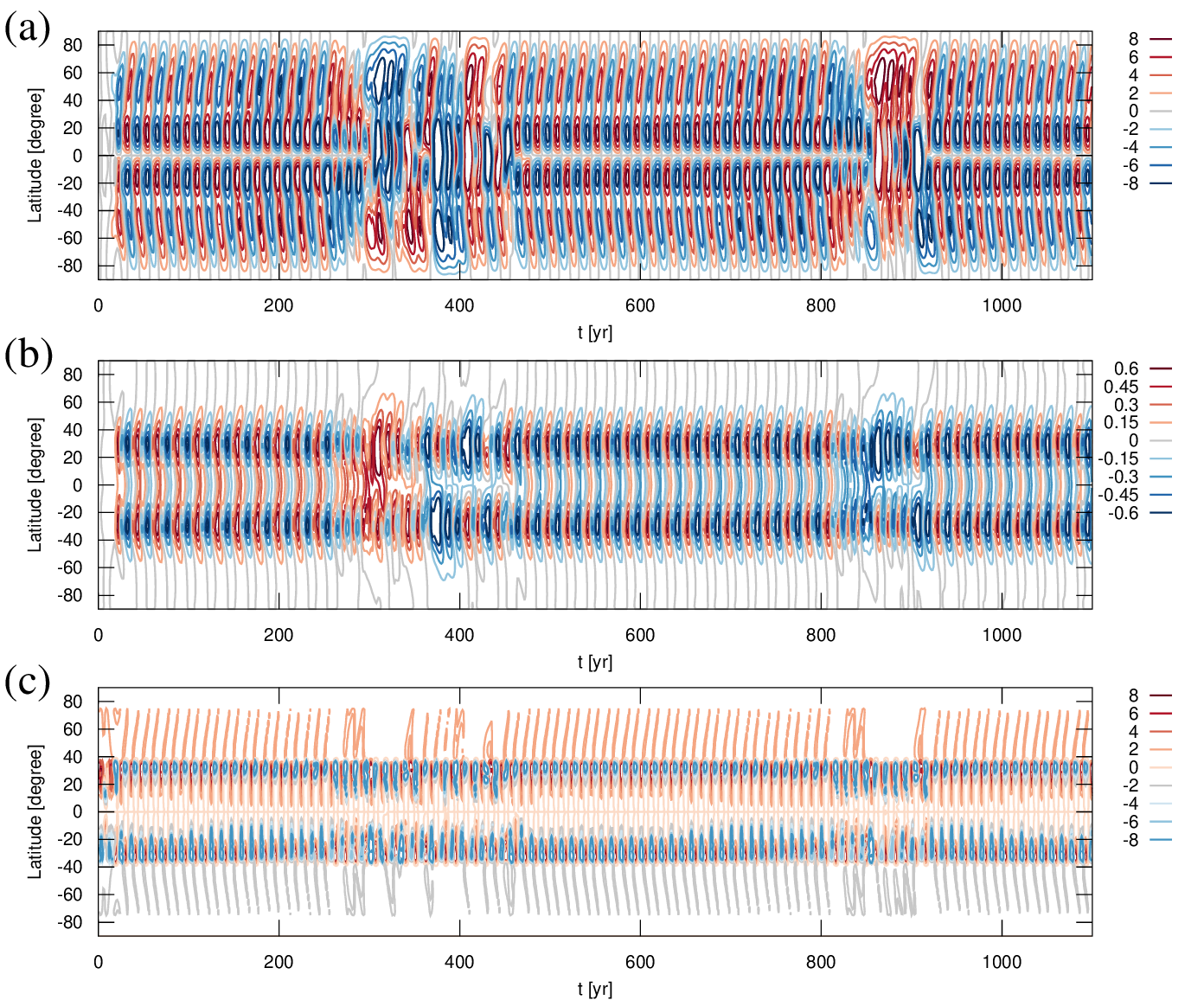}
\caption{Behaviour of $B(\theta,t)$ (a), $A(\theta,t)$ (b), 
and $\alpha(\theta,t)$ (c) showing transitions between  
dipole and quadrupole fields
when varying $\kappa$ according to $\kappa(t)=1 (1-0.6 \sin^2(2 \pi t/1100))$. The 
fixed parameters are $\Omega_0=10000$, $\alpha^c_0=4$, $\alpha^p_0(t)=100$
$q^p_{\alpha}=0.2$, $q^c_{\alpha}=0.8$.}
\label{Fig:maunder2}
\end{figure}

Figure \ref{Fig:maunder2} shows a similar result which we 
obtain when varying the loss term $\kappa B^3$ in Equation (1).
As noticed above, that term is supposed to account 
for the field losses due to 
magnetic buoyancy. Variations of this term might, therefore, be related
to variations of the adiabaticity, and hence of the
field storage capacity, in the tachocline, an effect 
that was proposed by \cite{Abreu2012} to explain 
the impact of weak tidal forces on 
(long-term) variations of the solar dynamo.
Again we see that these variations can lead to transitions between
dipoles and quadrupoles.
This means that, while only a synchronization 
of $\alpha$ seems to be strong enough to accomplish 
the ''hard synchronization'' 
of the basic Hale cycle, there is still 
a good chance that the long-term variations
of the solar cycle may also result from 
tidal effects on the adiabaticity in the tachocline.

\section{Discussion and outlook}

As a sequel to our previous studies \citep{Stefani2016,Stefani2018}, 
this paper was concerned with the spatio-temporal
behaviour of a tidally synchronized dynamo of the
Tayler-Spruit type, and its combination with a more
conventional $\alpha-\Omega$-dynamo. 
Utilizing a solar-like latitudinal dependence of the
$\Omega$-effect, and assuming a
plausible latitudinal dependence of the TI-related,
periodic  $\alpha$-effect, we have regularly found
dipole or quadrupole fields with 22.14 years periodic 
oscillations or
11.07 years periodic pulsations. Intermediate states between 
oscillations and
pulsations, reminiscent of the Gnevyshev-Ohl rule, as
well as hemispherical fields, were observed, too.
Under the influence of noise, the synchronized model 
maintained its character as a (globally) clocked process,
while a conventional $\alpha-\Omega$ model had  much closer
resemblance to a random walk process.

With appropriate changes of the relative weights of 
the periodic and the non-periodic
$\alpha$-terms, or by varying the loss term accounting for 
magnetic buoyancy, it
was easily possible to induce transitions between 
different field topologies, while maintaining  phase coherence
during all those transitions. The subcritical nature of the pure
Tayler-Spruit type model was confirmed, too.

The considered ``hybrid'' version of our synchronized dynamo, 
which builds 
on the conventional $\alpha-\Omega$ concept and requires only  weak 
periodic $\alpha$ forcing for synchronization, is quite attractive 
for the following reason: In the context of analyzing the two 
branches of main-sequence stars, separated by the Vaughan-Preston 
gap (around 2-3 Gyr, \cite{Vaughan1980}), the Sun appears as 
an ordinary, slowly rotating (older) star showing a typical 
activity period in the usual 10 years range, in contrast 
to faster rotating younger stars which 
show partly a shorter and strongly varying (7.6$\pm$4.9 years) 
periodicity, but in general a rather irregular temporal behavior 
\citep{Soon1993,Olah2016}. A bold explanation for the relation
between cycle period and rotation period, as observed for older stars, 
would have to assert that all of them were synchronized by a 
similar mechanism as discussed here. Since this scenario is rather 
unlikely (all those
stars would need planetary systems with a dominant tidal
periodicity in the same order of 10 years), 
we are in no way opposed to traditional dynamo concepts
yielding typical activity 
periods in the
order of 10 years. We suggest, however, that in particular 
cases such as our sun, those conventional dynamos 
could be synchronized by planetary tidal forcing.  
Our hybrid version
thus remedies the general fitting of our Sun into the
cycle period/rotation period relation of 
older stars
with the specific synchronization of the
Sun's dynamo as suggested by the time series 
of Figure 1 and the remarkable behaviour of 
Dicke's ratio shown in Figure 2. Unfortunately, similar 
statistical arguments as for the sun, 
which are based on tens or even hundreds of cycles, 
can not be inferred from 
the much shorter databases as available for 
other stars \citep{Soon1993,Olah2016}.

Two interesting features, which were already 
salient in the zero-dimensional model
of \cite{Stefani2016}, have been confirmed
in the 1D model: these are the appearance of a 
double peak of the field 
(best seen in the poloidal field), 
and the intermediate emergence of reversed helicities
in the two hemispheres. Both effects can indeed be 
related  to corresponding
observational facts.

Hence, our Tayler-Spruit type dynamo model, based on 
a tidally synchronized TI-related $\alpha$-effect,
might have acquired greater 
plausibility by evincing a number
of spatio-temporal features which are typical for the
solar magnetic field.  
We hope that these results are promising 
enough to  motivate more advanced 2D or 3D 
simulations.
It remains to be seen whether the evident weaknesses 
of the model, in particular the unconvincing
shape of butterfly diagram,
can be mitigated by such an advanced modelling. 
Just as more traditional concepts of the solar dynamo,
our model might still require an enhancement
by meridional circulation in order to show butterfly 
diagrams in their full beauty. It is here 
where also the specific role of the  
$\pm 55^{\circ}$ latitude region for starting the dynamo cycle
\citep{McIntosh2015} might find an explanation,
which could not be provided by our simple 1D model.

We would also point out that the main idea of our model, 
that the 
helicity of an $m=1$ instability can be synchronized even by 
a weak periodic $m=2$ tidal perturbation, with the energy
content of the instability being essentially unchanged, is not 
necessarily restricted to the very 
Tayler instability but might well
be applicable to other $m=1$ instabilities or 
flow features, too.
A preliminary study has shown, for example,
a comparable synchronization 
effect for the $m=1$ dominated Large Scale Circulation (LSC)
in Rayleigh-B\'enard convection \citep{Galindo}. 
Similar synchronization mechanisms 
have been discussed in connection with 
the $m=1$ eigenmode 
in the von-K\'{a}rm\'{a}n-sodium (VKS) dynamo experiment 
\citep{Giesecke2012,Giesecke2017}. It seems also 
worthwhile to examine the same 
$\alpha$ synchronization concept 
fot the recently discussed 
Rossby waves of the tachocline 
\citep{McIntosh2017,Dikpati2017,Zaqarashvili2018}.
The strong dependence of these waves 
on the gravity parameter would  bring back into play
the concept of a tidal influence 
on the adiabaticity as proposed by \cite{Abreu2012}.

Finally, we note that a completely 
new perspective for synchronization 
may arise from the 
recent observation that positive shear flows,
such as in the near-equator parts of the tachocline,
are susceptible to a new kind of axisymmetric, double-diffusive 
MRI, as long as both azimuthal and axial fields are present
\citep{Mama2018}.

\begin{acks}
This project has received funding 
from the European Research Council (ERC) under the 
European Union's Horizon 2020 research and innovation programm
(grant agreement No 787544).
The work was also supported in frame of the Helmholtz - RSF 
Joint Research Group ``Magnetohydrodynamic instabilities: Crucial 
relevance for large scale liquid metal batteries and
the sun-climate connection'', 
contract No HRSF-0044.
We would like to thank Norbert Weber for his numerical
work on the tidal synchronization of helicity 
oscillations.  Inspiring discussions with  
J\"urg Beer, Antonio Ferriz Mas, Peter Frick,
Laur\`{e}ne Jouve, 
G\"unther R\"udiger, Dmitry Sokoloff, Rodion Stepanov 
and Teimuraz Zaqarashvili 
on various aspects of the 
solar dynamo are gratefully acknowledged. We thank 
Willie Soon for pointing out the importance of
mid-term fluctuations, and for valuable comments 
on the sun-star connection problem. We highly appreciate the 
constructive criticism of the anonymous reviewer which 
prompted us to significantly revise the paper.
\end{acks}

\section*{Disclosure of Potential Conflicts of Interest}
The authors declare that they have no conflicts of interest.

\section*{Appendix}

\begin{figure}[h!]
\includegraphics[width=120mm]{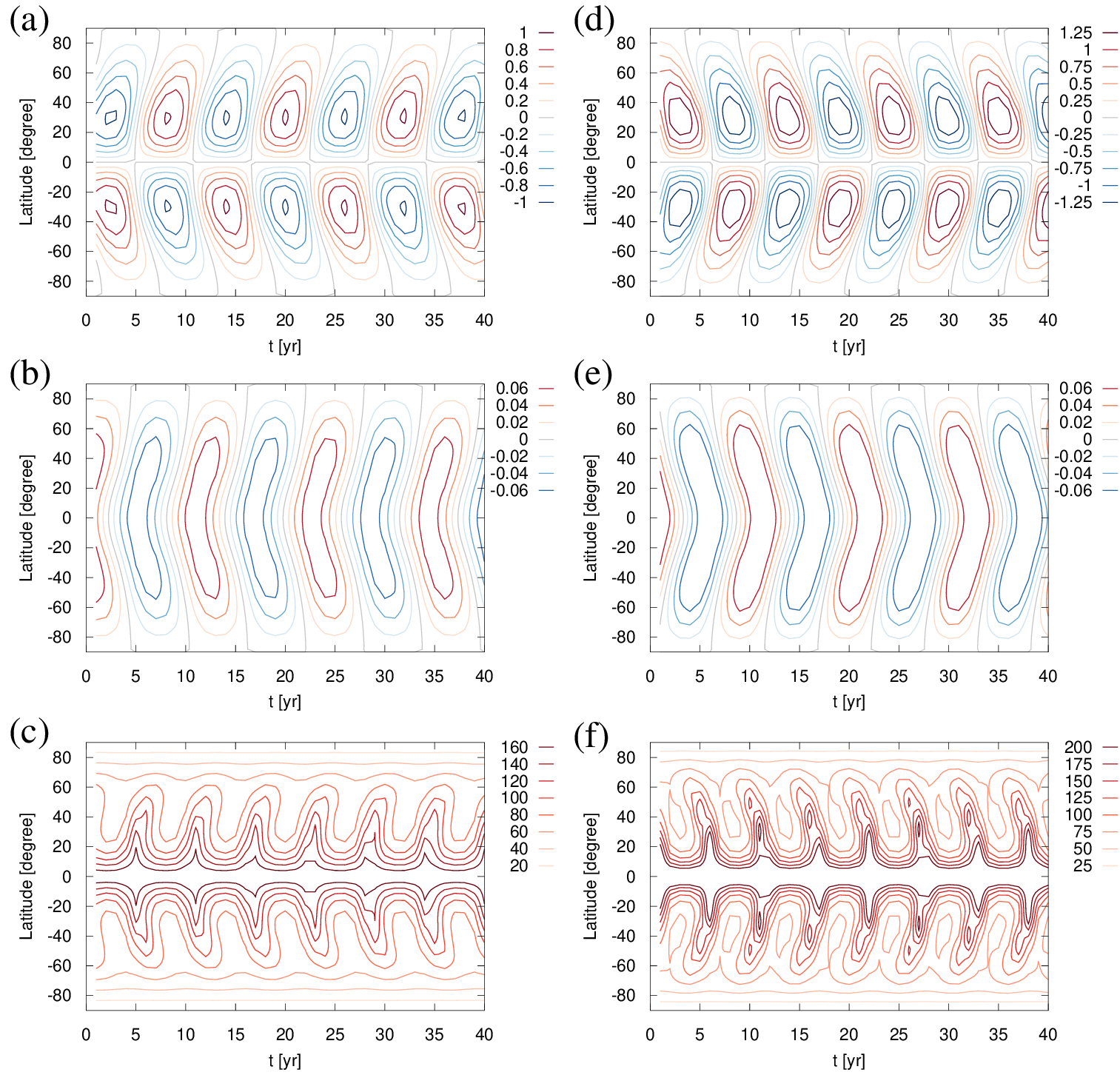}
\caption{Spatio-temporal behaviour of a simple $\alpha-\Omega$ model
with pure $\Omega$-quenching, for two different intensities
of the differential rotation, $\omega_0=170$ (a-c), and 
$\omega_0=250$ (d-f). The upper two panels (a,d) show $B(\theta,t)$,
the central two panels (b,e) show $A(\theta,t)$, the lower two panels 
(c,f) show  $\omega(\theta,t)$. Note the ''wrong'' butterfly direction for
$\omega_0=170$ (a), and the correct direction for $\omega_0=250$ (d).
In either case, the toroidal flux (a,d) is mainly transported along the
isolines of $\omega(\theta,t)$ (see c,f), according to Yoshimura's rule. 
}
\label{Fig:testrun}
\end{figure}

In this appendix, we validate our numerical model
by considering again the model of \cite{Jennings1991} which includes 
a (not very physical) quenching of the
$\Omega$-effect by the back-reaction of the magnetic 
field in the specific form
\begin{eqnarray}
\omega(\theta,t)&=&\omega_0 \sin(\theta)/(1+q_{\omega} B^2(\theta,t)) \; ,
\end{eqnarray}
while leaving the $\alpha$-effect unaffected.
Fixing $\alpha_0=-1$ and the quenching 
parameter $q_{\omega}=1$, 
Figure \ref{Fig:testrun} shows the
arising spatio-temporal dynamo behaviour for two different
values $\omega_0=170$ (a,b,c) and $\omega_0=250$
(d,e,f). The first row (a,d) shows $B(\theta,t)$, the second row
shows $A(\theta,t)$, and the third row shows 
 $\omega(\theta,t)$ (we skip 
 $\alpha(\theta)= \alpha_0 \cos(\theta)$ because it is 
 time-independent). 
 Interestingly, depending on the value
 of $\omega_0$, the system develops a butterfly diagram pointing either 
away from (a) or towards (d)  the equator. In either case, the direction 
follows basically the isolines of $\omega$, see (c) and (f),
according to a theorem by \cite{Yoshimura1975}.

\end{article} 

\end{document}